\newcommand{\beq}{\begin{equation}}
\newcommand{\eeq}{\end{equation}}
\newcommand{\bea}{\begin{eqnarray}}
\newcommand{\eea}{\end{eqnarray}}
\newcommand{\ee}{\epsilon}
\newcommand{\aaa}{\alpha}
\newcommand{\lll}{\lambda}
\newcommand{\vv}{\left\langle V\right\rangle_t}
\newcommand{\vvde}{\left\langle V^2\right\rangle_t}
\newcommand{\vvtr}{\left\langle V^3\right\rangle_t}
\newcommand{\vvs}{\left\langle V^s\right\rangle_t}
\newcommand{\vvr}{\left\langle V^r\right\rangle_t}
\newcommand{\xx}{\left\langle X\right\rangle_t}
\newcommand{\iii}{(i=1,2)}
\newcommand{\lla}{\left\langle}
\newcommand{\rra}{\right\rangle}
\newcommand{\noi}{\noindent}
\newcommand{\non}{\nonumber\\}
\newcommand{\mxw}{\stackrel{*}{=}}

\documentstyle[aps,epsf,floats]{revtex}

\begin{document}  

\title{On the adiabatic properties of a stochastic adiabatic wall: Evolution, 
stationary non-equilibrium, and equilibrium states}
\author{Ch. Gruber and L. Frachebourg}
\address{Institut de Physique Th\'eorique}
\address{Ecole Polytechnique F\'ed\'erale de Lausanne}
\address{CH-1015 Lausanne, Switzerland}
\date{\today} 
\maketitle

\begin{abstract}
 
The time evolution of the adiabatic piston problem and the consequences
of its stochastic motion are investigated.
The model is a one dimensional piston of mass $M$ separating two ideal
fluids made of point particles with mass $m\ll M$. For infinite systems it is
shown that the piston evolves very rapidly toward a stationary 
nonequilibrium state with non zero average velocity even
if the pressures are equal but the temperatures different on both sides of
the piston. 
For finite system it is shown that the evolution takes place in two stages:
first the system evolves rather rapidly and adiabatically toward a metastable
state where the pressures are equal but the temperatures different; then 
the evolution proceeds extremely slowly toward the equilibrium state 
where both the pressures and the temperatures are equal. Numerical simulations
of the model are presented. 
The results of the microscopical approach, the thermodynamical equations
and the simulations are shown to
be qualitatively in good agreement.  

\end{abstract}

\vskip 0.5truecm
\noi
{\bf Keywords:} Adiabatic piston; Boltzmann equation; Nonequilibrium 
thermodynamics.

\vskip 0.5truecm
\noi
{\bf PACS numbers} 05.20.Dd; 05.40.+j; 05.70.Ln; 02.50.Ey; 51.20.+d 

\vskip 1truecm

\section{Introduction}

\begin{figure}
\epsfxsize=10truecm
\hspace{3.25truecm}
\epsfbox{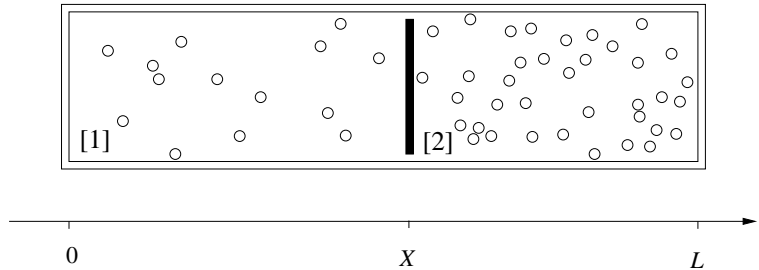}
\caption{
The adiabatic piston setting.
\label{fig1}}
\end{figure}

The {\sl adiabatic piston problem} is a well-known example in thermodynamics 
which has given rise to continuous controversy for the last 40 years 
(see \cite{gruber98} 
for selected references). It was also mentioned by E. Lieb 
as one of the
``problems in statistical mechanics that I would like to 
see solved'' at the StatPhys 20 meeting in 1998 \cite{lieb}. 
The problem is the 
following. The system is a finite cylinder containing two gases separated 
by an adiabatic movable piston. Initially, the piston is rigidly fixed
by a brake and the two gases are in thermal equilibrium characterized by
$\left(U_{[1]},V_{[1]},N_{[1]}\right)$ 
and $\left(U_{[2]},V_{[2]},N_{[2]}\right)$
(see Fig. \ref{fig1}).
At a certain time $t_0$, the brake is released 
and the question is to find the final equilibrium state.

Within the framework of {\sl thermostatics}, one only knows that the final 
equilibrium state
$\left(U_{[1]}(\infty),V_{[1]}(\infty),N_{[1]}(\infty)\right)$, 
$\left(U_{[2]}(\infty),V_{[2]}(\infty),N_{[2]}(\infty)\right)$, 
corresponds
to a maximum of the entropy function 
$S(U,V)=S_{[1]}+S_{[2]}=S_{[1]}(U,V)+S_{[2]}(U_0-U,V-V_0)$, where
$U_0=U_{[1]}+U_{[2]}$, $V_0=V_{[1]}+V_{[2]}$, submitted 
to the constraints $S_{[1]}\geq S_{[1]}(t_0)$, $S_{[2]}\geq S_{[2]}(t_0)$.
The solution of this mathematical problem yields the result that the final 
pressures must necessarily be equal $p_{[1]}(\infty)=p_{[2]}(\infty)$ 
({\sl i.e.} mechanical 
equilibrium), but nothing can be said about the final temperatures 
$T_{[1]}(\infty)$,
$T_{[2]}(\infty)$ (originally Kubo had arrived at the wrong conclusion 
$p_{[1]}(\infty)/T_{[1]}(\infty)=p_{[2]}(\infty)/T_{[2]}(\infty)$ 
\cite{kubo}). In other words the laws of
thermostatics are not sufficient to predict the final equilibrium state. Here
started the controversy, because some physicists could not 
accept this limitation of thermostatics. Callen writes in \cite{callen} 
``the movable adiabatic wall presents a unique problem with subtleties''.
The problem is however neither very subtle, 
nor unique since it appears every time 
movable adiabatic walls are present, and they are indeed essential in the
axiomatic formulation of thermostatics (see {\sl e.g.} \cite{lieb1}).

Recently, with the help of an oversimplified macroscopical model for the gases,
it was shown that the adiabatic piston problem, can be solved within the
framework of {\sl thermodynamics}\cite{gruber98}.
Equations for the time evolution were derived from the first and second law.
They show that the equilibrium state depends in an essential manner on the 
viscosity of the gases and one has to solve explicitly the time
evolution to find the final state,
which is however uniquely defined. In particular in the final state the 
temperatures $T_{[1]}(\infty)$ and $T_{[2]}(\infty)$ are different.

Using arguments from the {\sl kinetic theory of gases}, similar equations were
previously obtained by B. Crosignani et al. \cite{crosignani}, without the use 
of the laws of thermodynamics, but neglecting the effects of fluctuations.
The conclusion were however the same as the one obtained from thermodynamics,
namely $T_{[1]}(\infty)\ne T_{[2]}(\infty)$. 
At this point the controversy started again:
it was argued that the stochastic motion of the piston, induced by the
stochastic forces exerted by the particles of the gases, might lead to a
``true'' equilibrium state where $T_{[1]}(\infty)= T_{[2]}(\infty)$ 
and $p_{[1]}(\infty)=p_{[2]}(\infty)$.
In \cite{crosignani} it was stated ``this stochastic motion leads to
a violation of the second law'' and ``this would be equivalent to permit
heat to flow although the piston itself remains an insulator''. 
It was also suggested that this evolution toward the ``true''
equilibrium state, if it exists, might be so slow that it can never be 
observed in practice. 

The work done by J. L. Lebowitz in 1959 \cite{lebowitz} should have given
some insight into these questions already a long time ago. Indeed, using 
a model of non-interacting particles (in one dimension) to describe the gases,
 J. L. Lebowitz computed for the first time in \cite{lebowitz}, the heat 
conductivity of a movable ``adiabatic'' piston at lowest order in $\ee=
\sqrt{m/M}$ where $m$ is the mass of the gas particles and $M$ is the mass 
of the piston. However, we should remark that the model discussed in
\cite{lebowitz} was slightly different than the one considered here
since the piston was 
constrained to remain in mechanical equilibrium by means of an external force.

This same microscopical model, but without external force, was reintroduced by 
Piasecki and Gruber in \cite{gruber1}. Starting with the 
{\sl Boltzmann equation}, they obtained the exact stationary solution for
the {\sl infinite} one dimensional system, with $p_{[1]}(t)=
p_{[1]}=p_{[2]}=p_{[2]}(t)$, 
$T_{[1]}(t)=T_{[1]}\ne T_{[2]}=T_{[2]}(t)$, 
in the very special case $M=m$. The main conclusion
in \cite{gruber1} was that the space asymmetry ($T_{[1]}\ne T_{[2]}$) 
conspires
with the stochastic motion to give a stationary non-equilibrium state with
non-zero average velocity of the ``piston'' toward the region of the
higher temperature. The same model with $M=m$ was then considered by
Piasecki and Sinai\cite{sinai} to study the finite system. They were able
to show that the piston evolves toward an equilibrium position corresponding
to uniform density and uniform distribution of velocities (thus uniform 
``temperature''), for the particles throughout the system. 
Simultaneously Gruber and Piasecki investigated the more physical situation
$m\ll M$ \cite{gruber2}. 
Assuming that the solution of the Boltzmann equation has an 
asymptotic expansion in $\ee=\sqrt{m/M}$, the stationary solution of the 
{\sl infinite} system was computed at the order $\ee^2$. It was again 
established that the temperature difference ($T_{[1]}\ne T_{[2]}$) 
induces a
macroscopic motion toward the high temperature region despite the absence of 
any macroscopical force ($p_{[1]}=p_{[2]}$). In particular, it was shown 
that in this stationary state the piston behaves as a randomly moving particle
with kinetic energy $k_B\sqrt{T_{[1]}T_{[2]}}/2$, 
and average velocity ${\sqrt{m}\over M}\sqrt{{\pi k_B\over 8}}
\left(\sqrt{T_{[2]}}-\sqrt{T_{[1]}}\right)$.

In all these recent articles, the physical problems of heat conduction, 
viscous force, work power, as well as the approach toward the stationary,
or the equilibrium, state were not considered. These 
questions are investigated in the present article. 

Although we did not solve the problem mentioned by E. Lieb, which would need 
a rigorous mechanical many-body approach, the following analysis answers 
some of the conjectures discussed above and should contribute to a better 
understanding of the physical mechanisms involved in the time evolution
of the adiabatic piston. It raises also some conjectures that one might
be able to prove.
 
The main conclusion of this paper is that the phenomenological macroscopic
thermodynamics, the microscopical Boltzmann equation, and the numerical 
simulations, give results which co\"\i ncide with a very high degree
of accuracy when $m\ll M$. In other words there is no violation of the second 
law and no paradoxes involved.

The results of this investigation show that a wall which is ``adiabatic''
when rigidly fixed (no heat transfer), becomes ``conducting'' as soon as it 
is allowed to have a stochastic motion, and this is independent of the fact 
that its macroscopic velocity is zero or non-zero. This means that in this 
model heat transfer is associated with the stochastic motion of the piston, 
while work power is related to the macroscopic (or average) velocity of the 
piston $\vv$.
The time evolution obtained from thermodynamics 
(where the above conductivity is taken into account), 
from Boltzmann equation, and from 
the numerical simulations, shows that the system evolves toward a 
final state where the temperatures, as well as the pressures, are equal.
However, and this is the fundamental point, 
when $m\ll M$ the time scale needed
to reach this final state is so enormous (about the age of the universe 
for realistic numbers), that no reasonable person would call such a 
piston a conductor. More precisely our numerical analysis indicates 
that as soon as $m/M$ is small enough, 
the time evolution proceeds in two
stages with time scales which are very different. 
In the first stage the 
evolution is adiabatic and proceeds rather rapidly, with or without 
oscillations, towards a state of mechanical equilibrium where the pressures 
become equal, but
the temperatures remain different (this is the ``quasi-equilibrium state'' 
suggested in \cite{crosignani}). After this  quasi-equilibrium situation is 
reached the piston will slowly drift into the high temperature domain and
the system evolves, on a time scale several orders of magnitude 
larger, with equal pressures and slowly varying temperatures toward 
the final equilibrium state where temperatures, densities and pressures are 
equal on both sides of the piston.
On the other hand for 
$m/M\simeq 1$, the adiabatic piston is a good conductor and thermal
equilibrium $T_{[1]}(\infty)= T_{[2]}(\infty)$ is reached rapidly. 
Notice that on a very short time scale, there exists a stage where 
the piston behaves as if the system is infinite. In this stage, if the initial
pressures are different, the piston is accelerated while if the pressures are
initially equal the piston evolves with a (small) constant velocity inside the
high temperature fluid.

In the following analysis the microscopical model is a one dimensional 
system of point particles which do not interact except for purely
elastic collisions. It is exactly this model which has been simulated on 
a computer. 

For the analytical analysis we introduce three main assumptions. 
First we assume that the correlations between the velocities 
of the fluid particles and
the velocity of the piston can be neglected. 
Then we assume that 
the velocity distribution functions for the fluid particles on 
both faces of the piston admit a scaling function with respect to their 
mass $m$, as it is the case with Maxwellian distributions. 
However, because of the simulations (Sec.\ref{simul}), and the 
exact solution of Piasecki and Sinai \cite{sinai}, 
we do not assume Maxwellian
distribution for the velocity of the fluid particles. The temperatures 
$T_{[1]}(t)$ and $T_{[2]}(t)$ 
on both faces of the piston are then defined by the
second moment of the velocity of the fluid particles.
Finally we assume that the moments $\vvs$ of the velocity 
$V$ of the piston, $s\geq 1$,
have an asymptotic expansion in powers of $\ee=\sqrt{m/M}$.

To discuss the solutions of the time evolution equations, we shall 
either assume that the fluids densities and temperatures 
are constant (in the case of
an infinite system), or uniform (in the case of a finite system).

The microscopical model 
is defined in Sec.\ref{micro}. In Sec.\ref{conserve} we 
use the conservation laws of the problem to derive the dynamical
equations for the
first and second moment of the piston velocity distribution and give in
Sec.\ref{arbitrary} these equations for arbitrary moments $\vvs$.
The relation between the microscopical model and a phenomenological
thermodynamic approach is discussed in Sec.\ref{pheno} where we derive
a microscopical definition of the friction coefficients 
and the heat conductivity. The analysis
of the stationary non-equilibrium, and the equilibrium, states is presented in
Sec.\ref{stationary}, while the time evolution is discussed in 
Sec.\ref{timeevol}.
The numerical simulations, the comparison with the dynamic equations, and the
thermodynamic equations, is the subject of Sec.\ref{simul}. Finally, 
conclusions are given in Sec.\ref{conclusion}.

\section{Microscopical model}\label{micro}

We consider the fluids to be perfect gases made of $N_{[1]}$, respectively 
$N_{[2]}$, 
identical non-interacting 
point particles of mass $m$, making purely elastic collisions, and 
contained in a fixed cylinder of length $L$.  The basis of the cylinder is a 
circle with area $A$. The adiabatic piston is a rigid solid ({\sl i.e.} 
no internal degree of freedom), with mass $M$, described by the microscopical 
variables $(X,V={dX\over dt})$ (see Fig. \ref{fig1}).

We assume that the cylinder is an ideal constraint, which has no physical
 properties: the collisions of the particles on the cylinder are purely 
elastic, and there is no friction between the piston and the cylinder.

For $t\leq 0$, the piston is fixed $(X=X_0,V=0)$ and both fluids are in  
thermal equilibrium, with temperatures $T_{[1]}$, 
respectively  $T_{[2]}$, 
described by the Maxwellian distribution functions:
\bea
&& \rho_{[1]}(x,v;t\leq 0)=\rho_{[1]}\phi_{T_{[1]}}(v)\quad{\rm for} 
\quad 0\leq x< X_0
\nonumber\\
&& \rho_{[2]}(x,v;t\leq 0)=\rho_{[2]}\phi_{T_{[2]}}(v)\quad{\rm for} 
\quad X_0< x\leq L
\label{1}
\eea
where 
\beq
\rho_{[1]}={1\over A}{N_{[1]}\over  X_0}={1\over A}n_{[1]},\quad 
\rho_{[2]}={1\over A}{N_{[2]}\over L-X_0}={1\over A}n_{[2]}
\eeq
and
\beq
\phi_{T}(v)=
\sqrt{{m\over 2\pi k_BT}}\exp\left(-{mv^2\over 2k_BT}\right).
\label{max}
\eeq

At time $t=0$, the break is released and the piston can move freely. It will 
thus have a stochastic motion induced by collisions with the fluids
and $(X,V)$ become random variables. 
The problem is
to investigate the random motion of the piston.
To simplify the discussion, we consider the system to be one-dimensional. 
In this case the adiabatic piston is just a point particle with mass $M$.

For $t>0$, the piston moves freely. Since the collisions are assumed to be 
purely elastic, then if the masses $M$ and $m$ have velocities $V$ and $v$ 
before the collision, their velocities after the collision are given by
\bea
&& v'=v-{2M\over M+m}(v-V)\nonumber\\
&& V'=V+{2m\over M+m}(v-V)
\label{coll}
\eea

The main assumption of our analysis is that the time evolution is such 
that it is possible to neglect the correlations between the velocity of 
the piston and the velocities  
of the fluids particles. This means that we assume 
that the joint probability distribution to find, at time $t$, 
the piston at position $X$ with velocity $V$ and a fluid particle at position 
$X\pm |\ee|$ with velocity $v$ can be expressed by 
\bea
\Phi_{[1]}(X,V,X,v;t)&=&\Phi(X,V;t)\lim_{\ee\to 0}\rho_{[1]}(X-|\ee|,v;t)=
\Phi(X,V;t)n_{[1]}(t)\phi_{[1]}(v,t)\non
\Phi_{[2]}(X,V,X,v;t)&=&\Phi(X,V;t)\lim_{\ee\to 0}\rho_{[2]}(X+|\ee|,v;t)=
\Phi(X,V;t)n_{[2]}(t)\phi_{[2]}(v,t)
\label{assump}
\eea
where we have introduced the time dependent 
particle densities on the faces of the
piston $n_{[i]}$ and the normalized fluids velocity 
distribution functions on the faces of the piston $\phi_{[i]}(v,t)$ 
for $i=1,2$.

In the case of an infinite cylinder, we shall consider that 
$n_{[1]}(t)$ and $n_{[2]}(t)$ are constant, equal to 
the fluid initial densities, 
and that $\phi_{[i]}(v,t)=\phi_{T_{[i]}}(v)$ ($i=1,2$). 
This assumption is justified if $m\ll M$, since 
in this case there is a vanishing
probability that the piston interact back with the perturbation it causes in 
the state of the surrounding fluids (no recollision). In other words, the 
piston always ``sees'' on both sides the unperturbed initial states.

For the finite system, when $m\ll M$, we expect the average velocity 
of the piston to be small, and therefore the time needed for a macroscopical
but small displacement of the piston will be much larger than the relaxation 
time of the fluids 
(if the initial pressures difference is not too large).
We shall then assume in Sec.\ref{arbitrary} 
that at all time $t$, the 
fluids on both side of the piston are homogeneous and thus that
the distribution functions $\rho_{[1]}(x,v,t)$, $0<x<X$, and
$\rho_{[2]}(x,v,t)$, $X<x<L$, are independent from the
position $x$. We then have 
\beq
n_{[1]}(t)={N_{[1]}\over \left\langle X\right\rangle_t}, \quad
n_{[2]}(t)={N_{[2]}\over L-\left\langle X\right\rangle_t}
\eeq
and the time-dependent temperatures $T_{[1]}(t)$ and $T_{[2]}(t)$ are 
defined by the second moment of the fluid 
velocity.

For the following analysis, we introduce the small parameters 
\beq
\alpha={2m\over M+m}\quad {\rm and}\quad \epsilon=\sqrt{{m\over M}}.
\eeq

\section{Conservation laws}\label{conserve}

We use the conservation laws present in the problem to determine the dynamical 
equations
for moments of the piston velocity distribution. 
In particular, the conservation
of momentum and energy leads to equations for the first and second moment of 
the piston velocity distribution. 

\subsection{Linear momentum}

Let $\Pi_{[1]}(t)$ and $\Pi_{[2]}(t)$ denote the 
linear momentum of the fluids on 
the left and on the right of the piston at time $t$. Using the collision 
equations (\ref{coll}) and the assumption (\ref{assump}) the force
exerted by the left fluid on the piston is
\beq
F^{[1]\to p}=-\lim_{\delta t\to 0}
\left\langle {\delta \Pi_{[1]}(t)\over \delta t}\right\rangle=
M\alpha\int_{-\infty}^\infty dV \Phi_\ee(V;t)\int_V^\infty dv\, 
n_{[1]}(t)\phi_{[1]}(v,t)(v-V)^2
\label{8}\eeq
where $\Phi_\ee(V;t)$ is the time dependent 
velocity distribution function of the piston.
Similarly the force exerted by the right fluid on the piston is
\beq
F^{[2]\to p}=
-\lim_{\delta t\to 0}\left\langle {\delta \Pi_{[2]}(t)\over \delta t}
\right\rangle=
-M\alpha\int_{-\infty}^\infty dV \Phi_\ee(V;t)\int_{-\infty}^V dv\, 
n_{[2]}(t)\phi_{[2]}(v,t)(v-V)^2.
\label{9}
\eeq
Therefore, the conservation law of linear momentum
\beq
M{d\over dt}\vv-F^{[1]\to p}-F^{[2]\to p}=
M{d\over dt}\vv+ \lim_{\delta t\to 0}\left[
\left\langle {\delta \Pi_{[1]}(t)\over \delta t}
\right\rangle+\left\langle {\delta \Pi_{[2]}(t)\over \delta t}
\right\rangle\right]=0
\label{10}
\eeq
can be written in the following form
\bea
{d\over dt}\vv &=& \alpha
\int_{-\infty}^\infty dV \Phi_\ee(V;t)\int_0^\infty dv\, 
\left[n_{[1]}(t)\phi_{[1]}(v,t)(v-V)^2-n_{[2]}(t)\phi_{[2]}(-v,t)(v+V)^2
\right] \nonumber\\
&& \quad -\alpha \int_{-\infty}^\infty dV \Phi_\ee(V;t)\int_0^V dv\,(v-V)^2
\left[n_{[1]}(t)\phi_{[1]}(v,t)+n_{[2]}(t)\phi_{[2]}(v,t)\right].
\label{mom0}
\eea

Using  the expansion
\beq
\phi_{[i]}(v,t)=\sum_{k=0}^\infty 
{v^k\over k!}\phi_{[i]}^{\prime(k)}(t),\quad (i=1,2)
\label{expansion}
\eeq
where
\beq
\phi^{\prime(k)}_{[i]}(t) 
= \left.\left({d\over dv}\right)^k\phi_{[i]}(v,t)\right|_{v=0},
\quad (i=1,2),
\eeq
we obtain
\bea
{d\over dt}\vv &=& \alpha\left[n_{[1]}(t)\left\langle v^2\right\rangle^{[1]}_t-
n_{[2]}(t)\left\langle v^2\right\rangle^{[2]}_t\right]\non
&&\quad-\vv 2\alpha \left[n_{[1]}(t)\left\langle v\right\rangle^{[1]}_t+
n_{[2]}(t)\left\langle v\right\rangle^{[2]}_t\right]\nonumber \\
&&\quad\quad  
+\left\langle V^2\right\rangle_t \alpha \left[n_{[1]}(t)\left\langle 
v^0\right\rangle^{[1]}_t
-n_{[2]}(t)\left\langle v^0\right\rangle^{[2]}_t\right]\non
&&\quad\quad\quad
-\sum_{k=0}^\infty \left\langle V^{k+3}\right\rangle_t{2\alpha\over (k+3)!}
\left[n_{[1]}(t)\phi_{[1]}^{\prime(k)}(t)
+n_{[2]}(t)\phi_{[2]}^{\prime(k)}(t)\right]
\label{mom1}
\eea
where for any non negative integers $s$ 
\bea
\left\langle v^s\right\rangle^{[1]}_t 
&=& \int_0^\infty dv\,\phi_{[1]}(v,t)v^s,
\nonumber\\
\left\langle v^s\right\rangle^{[2]}_t  
&=& \int_0^\infty dv\,\phi_{[2]}(-v,t)v^s,\quad {\rm and}
\nonumber\\
\vvs &=& 
\int_{-\infty}^{\infty} dV\, \Phi_\ee(V;t)V^s.
\label{phi0}
\eea
We should insist on the fact that in the definitions of 
$\left\langle v^s\right\rangle^{[i]}_t$ the integration is on $v$ positive.
In particular $\left\langle v^0\right\rangle^{[i]}_t$ is not equal to 1.

We should note that the evolution equation (\ref{mom1}) for 
the average velocity of the
piston depends on higher moments of the piston velocity.
We should also remark that if the term 
$n_{[1]}(t)\lla v^2\rra^{[1]}_t-n_{[2]}(t)\lla v^2\rra^{[2]}_t$ is zero 
(which means, as we shall see in Sec.\ref{pheno}, that the 
pressures exerted by the fluids are equal) and
if we neglect the fluctuations of the piston velocity 
($\vvs=\vv^s$) we find the stable equilibrium solution $\vv=0$. 
The effects of the fluctuations is the main problem to be investigated and 
will give a very different picture.

\subsection{Energy}

Similarly, let $E_{[1]}(t)$ and $E_{[2]}(t)$ denote the energy 
of the fluids on 
the left and on the right of the piston at time $t$. 
It is equal to the kinetic energy, since the particles do not interact
except for elastic collisions. Using the collision equations (\ref{coll}) 
and the assumption (\ref{assump}),
we have
\beq
\lim_{\delta t\to 0}\left\langle 
{\delta E_{[1]}(t)\over \delta t}\right\rangle=
-M\alpha\int_{-\infty}^\infty dV \Phi_\ee(V;t)\int_V^\infty dv\, 
n_{[1]}(t)\phi_{[1]}(v,t)(v-V)^2
\left[V+{\alpha\over 2}(v-V)\right]
\label{16}
\eeq
and
\beq
\lim_{\delta t\to 0}\left\langle 
{\delta E_{[2]}(t)\over \delta t}\right\rangle=
M\alpha\int_{-\infty}^\infty dV \Phi_\ee(V;t)\int_{-\infty}^V dv\, 
n_{[2]}(t)\phi_{[2]}(v,t)(v-V)^2
\left[V+{\alpha\over 2}(v-V)\right].
\eeq
Therefore, the conservation of energy
\beq
{1\over 2}M{d\over dt}\left\langle V^2\right\rangle_t+
\lim_{\delta t\to 0}\left[
\left\langle {\delta E_{[1]}(t)\over \delta t}\right\rangle+
\left\langle {\delta E_{[2]}(t)\over \delta t}\right\rangle\right]=0
\label{17}
\eeq
leads to
\bea
&&{d\over dt}\vvde =
2\alpha
\int_{-\infty}^\infty dV \Phi_\ee(V;t)\int_0^\infty dv\, \non
&&\quad\quad
\left\{n_{[1]}(t)\phi_{[1]}(v,t)(v-V)^2\left[V+{\alpha\over 2}(v-V)\right]
-n_{[2]}(t)\phi_{[2]}(-v,t)(v+V)^2\left[V-{\alpha\over 2}(v+V)\right]
\right\} \nonumber\\
&& \quad -2\alpha \int_{-\infty}^\infty dV 
\Phi_\ee(V;t)\int_0^V dv\,(v-V)^2\left[V+
{\alpha\over 2}(v-V)\right]
\left[n_{[1]}(t)\phi_{[1]}(v,t)+n_{[2]}(t)\phi_{[2]}(v,t)\right].
\eea
With the expansion (\ref{expansion}) and the definitions (\ref{phi0}) 
this last expression can be written as
\bea
{d\over dt}\vvde &=& \alpha^2\left[n_{[1]}(t)
\left\langle v^3\right\rangle^{[1]}_t+
            n_{[2]}(t)\left\langle v^3\right\rangle^{[2]}_t\right]\nonumber\\
&& +\vv\aaa(2-3\aaa)\left[n_{[1]}(t)\left\langle v^2\right\rangle^{[1]}_t-
        n_{[2]}(t)\left\langle v^2\right\rangle^{[2]}_t\right]\nonumber\\
&& -\vvde\aaa(4-3\aaa)\left[n_{[1]}(t)\left\langle v\right\rangle^{[1]}_t+
           n_{[2]}(t)\left\langle v\right\rangle^{[2]}_t\right]\nonumber\\
&& +\vvtr\aaa(2-\aaa)\left[n_{[1]}(t)\left\langle v^0\right\rangle^{[1]}_t-
            n_{[2]}(t)\left\langle v^0\right\rangle^{[2]}_t\right]\nonumber\\
&& -\sum_{k=0}^\infty \left\langle V^{k+4}\right\rangle_t{2\alpha\over (k+4)!}
(2k+8-3\aaa)\left[n_{[1]}(t)\phi_{[1]}^{\prime(k)}(t)
+n_{[2]}(t)\phi_{[2]}^{\prime(k)}(t)\right]
\label{mom2}
\eea

We thus obtained an equation for the evolution of the second moment of the 
piston velocity. 
Let us remark that even in case of mechanical equilibrium where 
$n_{[1]}(t)\left\langle v^2\right\rangle^{[1]}_t-
n_{[2]}(t)\left\langle v^2\right\rangle^{[2]}_t=0$ 
the initial state $\left\langle V^s\right\rangle_0=0$ 
is unstable if 
$n_{[1]}(t)\left\langle v^3\right\rangle^{[1]}_t
+n_{[2]}(t)\left\langle v^3\right\rangle^{[2]}_t\ne 0$.
In conclusion, 
if we do not neglect the correlations in (\ref{mom1}), we see that
even in case of mechanical equilibrium,
the initial state $\left\langle V^s\right\rangle_0=0$ is unstable 
and thus the piston will start to move.

\section{Equation for arbitrary moment}\label{arbitrary}

In the previous section, we derived equations for the first and second 
moments of the piston velocity which are coupled with higher order moments.
We derive in this section  
the infinite set of coupled equations for all the moments 
of the piston velocity.

We use the Boltzmann's equation for the piston velocity
distribution derived in \cite{gruber1} under the assumption (\ref{assump})
\bea
{\partial \over \partial t}\Phi_\ee(V;t)&=& 
\int_{-\infty}^\infty dv\, |V-v|\left[
\Theta(V-v) n_{[1]}(t)\phi_{[1]}\left(v-(2-\aaa)(v-V);t\right)
\Phi_\ee\left(V+\aaa(v-V);t\right)\right.\nonumber\\
&& \quad\left.
+\Theta(v-V)n_{[2]}(t)\phi_{[2]}\left(v-(2-\aaa)(v-V);t\right)
\Phi_\ee\left(V+\aaa(v-V);t\right)\right.\nonumber\\
&& \quad\left.-\Theta(v-V)n_{[1]}(t)\phi_{[1]}(v,t)\Phi_\ee(V,t)
-\Theta(V-v)n_{[2]}(t)\phi_{[2]}(v,t)\Phi_\ee(V,t)\right]
\eea
to obtain the equations for the velocity moments 
\bea
{d\over dt}\vvs &=& \int_{-\infty}^\infty dV V^s {\partial \over \partial t}
\Phi_\ee(V;t)
\nonumber\\
&=& \left\langle \int_V^\infty dv\, n_{[1]}(t)\phi_{[1]}(v,t)(v-V)\left\{\left[
V(1-\aaa)+\aaa v\right]^s-V^s\right\}\right\rangle_t\nonumber\\
&& \quad 
-\left\langle \int_{-\infty}^V dv\, n_{[2]}(t)\phi_{[2]}(v,t)(v-V)\left\{\left[
V(1-\aaa)+\aaa v\right]^s-V^s\right\}\right\rangle_t
\eea
where $\left\langle A(V)\right\rangle_t=\int_{-\infty}^\infty dV A(V) 
\Phi_\ee(V;t)$.
This equation can be written in the following form
\bea
&&{d\over dt}\vvs = 
\left\langle \int_0^\infty dv\, n_{[1]}(t)\phi_{[1]}(v,t)(v-V)\left\{\left[
V(1-\aaa)+\aaa v\right]^s-V^s\right\}\right\rangle_t\nonumber\\
&& \quad\quad +\left\langle \int_0^\infty dv\, n_{[2]}(t)
\phi_{[2]}(-v,t)(v+V)\left\{\left[
V(1-\aaa)-\aaa v\right]^s-V^s\right\}\right\rangle_t\nonumber\\
&& \quad\quad -\left\langle \int_0^V dv\, (v-V)\left\{\left[
V(1-\aaa)+\aaa v\right]^s-V^s\right\}
\left[ n_{[1]}(t)\phi_{[1]}(v,t)+  
n_{[2]}(t)\phi_{[2]}(v,t)\right]\right\rangle_t.
\eea
Using the expansions (\ref{expansion}) and 
\beq
\left[V+\aaa (v-V)\right]^s-V^s=\sum_{q=1}^{s}{s!\over q!(s-q)!}
\aaa^q(v-V)^qV^{s-q},
\eeq
we obtain
\beq
{d\over dt}\vvs=\sum_{r=0}^\infty A_{s,r}(t)\left\langle V^r\right\rangle_t
\label{serie0}
\eeq
where one has the explicit expression of $A_{s,r}(t)$ as a function of
the parameter $\aaa$. In the following, it will be more convenient to have
the explicit expression of $A_{s,r}(t)$ as a function of the parameter
$\ee=\sqrt{m/M}$. 

Motivated by the Maxwellian distribution (\ref{max}), we
introduce the scaled velocity $w=\sqrt{m}v$ 
and the corresponding normalized distribution functions
$f_{[i]}(w,t)$
\beq
\phi_{[i]}(v,t)=\sqrt{m}f_{[i]}(\sqrt{m} v,t),\quad \int_{-\infty}^\infty dw\, 
f_{[i]}(w,t)=1,\quad (i=1,2).
\label{scale}
\eeq
We then have from Eqs.~(\ref{phi0},\ref{scale}),
\beq
\left\langle v^q\right\rangle_t^{[i]}=m^{-q/2}\int_0^\infty dw f_{[i]}(w;t) 
w^q= m^{-q/2} \left\langle w^q\right\rangle_t^{[i]},\,\,(i=1,2)
\eeq
and
\beq
\phi_{[i]}^{\prime(k)}(t)=m^{(k+1)/2}f_{[i]}^{\prime(k)}(t).
\eeq
For any integer $r$ we introduce the functions $K_r^{[i]}(t)$ and
$K_r(t)$ defined in the following manner. 
For any $q\geq 0$, and $i=1,2$, 
\bea
K_{-2+q}^{[i]}(t)&=& M^{-q/2}n_{[i]}(t)\left\langle  
w^{q}\right\rangle_t^{[i]},\\
K_{-3-q}^{[i]}(t)&=& M^{(q+1)/2}n_{[i]}(t) f_{[i]}^{\prime(q)}(t),\\
K_r(t)&=& K_r^{[1]}(t)-(-1)^r K_r^{[2]}(t)\quad{\rm for}\,\, r\geq -2,
\label{kr1}\\
K_r(t)&=& K_r^{[1]}(t)+K_r^{[2]}(t)\quad{\rm for}\,\, r\leq -3.
\label{kr2}
\eea
With these definitions, we have
\beq
{d\over dt}\vvs=\sum_{r=0}^\infty A_{s,r}(t)\left\langle V^r\right\rangle_t
\label{serie}
\eeq
where the coefficients $A_{s,r}(t)$ are given by
\beq
A_{s,r}(t)={\ee^{|s-r-1|}\over (1+\ee^2)^{s}}P_{s,r}(\ee^2) K_{s-r-1}(t).
\label{34}
\eeq
The functions $K_r(t)$, Eqs.~(\ref{kr1},\ref{kr2}), 
depend only on the densities 
and the velocity distribution functions of the fluid particles 
on both faces of the piston, while $P_{s,r}(\ee^2)$ is a polynomial in
$\ee^2$, of order $\min(r,s-1)$, with constant coefficients, given by
\beq
P_{s,r}(\ee^2)=
(1-\ee^2)^{r-1}\left[(s+1-r)
-\ee^2(s+1+r)\right] {2^{s-r} s!\over r! (s-r+1)!}, \quad r\leq s-1;
\eeq
\bea
P_{s,s}(\ee^2)&=& {(1-\ee^2)^{s-1}\left[1-\ee^2(2s+1)
\right]-(1+\ee^2)^s\over \ee^2};\\
P_{s,s+1}(\ee^2)&=& {(1+\ee^2)^{s}-(1-\ee^2)^s
\over \ee^2};
\eea
\beq
P_{s,s+2+k}(\ee^2) = -2\sum_{q=0}^{s-1}(-1)^q (2\ee^2)^q(1+\ee^2)^{s-q-1}
{s!(q+2)\over (s-1-q)!(k+q+3)!}\quad r\geq s+2.
\eeq
One can verify that Eq.(\ref{serie}) for $s=1,2$ are identical to the 
equations for the first and the second moment $(\ref{mom1},\ref{mom2})$ 
derived in the preceding section.

\section{From kinetic theory to thermodynamics: 
Friction coefficient and heat conduction}\label{pheno}

In this section we relate the microscopical model to a phenomenological 
thermodynamic approach. This will suggest 
microscopical definitions for ``pressure'', ``viscous force'', ``temperature''
and ``heat flux''.

Using a very primitive model for the fluids, the equations for the time 
evolution of the piston problem were derived in \cite{gruber98} 
from the first and second laws of the thermodynamics. They read
\bea
(M+\delta M) {dV\over dt} &=& A(p_{[1]}-p_{[2]})
-(\lll_{[1]}+\lll_{[2]})V;\quad V={dX\over dt}
\label{eqv}\\
{dS_{[1]}\over dt}&=&{\lll_{[1]}\over T_{[1]}}V^2
+{\kappa\over T_{[1]}}(T_{[2]}-T_{[1]})
\label{eqs1}\\
{dS_{[2]}\over dt}&=&{\lll_{[2]}\over T_{[2]}}V^2
-{\kappa\over T_{[2]}}(T_{[2]}-T_{[1]})
\label{eqs2}\eea
where $X$ and $V$ are the macroscopic position and 
velocity of the piston of mass $M$, 
$\delta MV^2/2$ represents the kinetic energy of the fluids, $A$ is the area 
of the cylinder, 
$p_{[1]},p_{[2]},\lll_{[1]},\lll_{[2]},\kappa$ are phenomenological
time-dependent functions of the variables 
$(X,V,S_{[1]},S_{[2]})$ and denote the pressures, the friction coefficients, 
and the heat conductivity. 
$S_{[1]}$ and $S_{[2]}$ denote the entropy of the fluids on 
the left and on the right 
of the piston.

For ideal fluids in one dimension, the thermal equation (\ref{eqs1}) and 
(\ref{eqs2}) are equivalent to
\bea
{1\over 2} N_{[1]} k_B {dT_{[1]}\over dt} &=& \lll_{[1]} V^2 
-{N_{[1]} k_B T_{[1]}\over X}V
+\kappa(T_{[2]}-T_{[1]}) \label{42}\\
{1\over 2} N_{[2]} k_B {dT_{[2]}\over dt} &=& \lll_{[2]} V^2 
+{N_{[2]} k_B T_{[2]}\over L-X}V
-\kappa(T_{[2]}-T_{[1]})\label{43}.
\eea

In thermodynamics, the adiabatic piston is defined by 
$\kappa=0$. One should remark that in this case 
($\kappa=0$), 
if the velocity of the piston is sufficiently small so that the term in $V^2$ 
can be neglected, the evolution of the piston is damped, but
reversible in the thermodynamical sense, characterized by the adiabats
\bea
S_{[1]}(t) = S_{[1]} \quad &{\rm and}& \quad S_{[2]}(t) = S_{[2]}\nonumber\\
{\sl i.e.}\quad\quad\quad
T_{[1]}X^2 = {\rm cte}  \quad && \quad T_{[2]}(L-X)^2 = {\rm cte}.
\eea
In particular, the solution $p_{[1]}=p_{[2]}$, $V=0$ is stable.

To make the connection with
the thermodynamical equations (\ref{eqv}), we write the microscopic 
equation for conservation of momentum (\ref{mom0}) or (\ref{mom1}) in the
form
\bea
&& M\left(1+{m\over M}\right){d\over dt}\vv =
2m\left[n_{[1]}(t)\left\langle v^2\right\rangle^{[1]}_t-
n_{[2]}(t)\left\langle v^2\right\rangle^{[2]}_t\right]\nonumber \\
&&\quad -\vv 2m n_{[1]}(t) \left[2\left\langle v\right\rangle^{[1]}_t
-\vv\left\langle v^0\right\rangle^{[1]}_t
+{1\over \vv}\int_0^{\vv} dv\,\phi_{[1]}(v,t)(v-\vv)^2\right]\nonumber \\
&&\quad -\vv 2m n_{[2]}(t) \left[2\left\langle v\right\rangle^{[2]}_t
+\vv\left\langle v^0\right\rangle^{[2]}_t
+{1\over \vv}\int_0^{\vv} dv\,\phi_{[2]}(v,t)(v-\vv)^2\right]\nonumber \\
&&\quad +\left[\left\langle V^2\right\rangle_t
-\left\langle V\right\rangle_t^2\right]
 2m \left[n_{[1]}(t)\left\langle v^0\right\rangle^{[1]}_t-
n_{[2]}(t)\left\langle v^0\right\rangle^{[2]}_t\right]\nonumber \\
&&\quad -\sum_{k=0}^\infty \left[\left\langle V^{k+3}\right\rangle_t-
\left\langle V\right\rangle_t^{k+3}\right]{4m\over (k+3)!}
\left[n_{[1]}(t)\phi_{[1]}^{\prime(k)}(t)
+n_{[2]}(t)\phi_{[2]}^{\prime(k)}(t)\right].
\label{mom1bis}
\eea
We are thus led to identify $\vv$ with the macroscopic velocity $V(t)$ 
of the piston, and to define the ``pressure'', the ``temperature'', and the
``friction coefficient'', on both sides ($i=1,2$) of the 
piston by 
\beq
p_{[i]}(t)=n_{[i]}(t)k_BT_{[i]}(t)=2mn_{[i]}(t)
\left\langle v^2\right\rangle^{[i]}_t=2n_{[i]}(t)
\left\langle w^2\right\rangle^{[i]}_t=2MK_0^{[i]}(t)
\label{tempdefine}
\eeq
\beq
\lll_{[i]}(V,t)=2mn_{[i]}(t)\left[2\left\langle v\right\rangle^{[i]}_t
+(-1)^i V\left\langle v^0\right\rangle^{[i]}_t
+{1\over V}\int_0^V dv\,\phi_{[i]}(v,t)(v-V)^2\right]
\label{defcoef}
\eeq
which gives in particular
\beq
\lll_{[i]}(V=0)=4\sqrt{m}n_{[i]}\left\langle w\right\rangle^{[i]}
=4\ee MK_{-1}^{[i]},
\label{lllv0}
\eeq
\beq
\lll_{[1]}(V)+\lll_{[2]}(V)=M\left[4\ee K_{-1}
-2\ee^2K_{-2}V+{\cal O}(\ee^3 V^2))
\right],
\eeq
and shows that the condition $\lll_{[i]}(V)\geq 0$ will be satisfied in the 
domain considered here where the average piston velocity is 
much smaller than the average absolute velocity of a fluid particle,
$V\ll\left\langle v\right\rangle^{[i]}_t$. Notice that the temperature 
definition Eq.(\ref{tempdefine}) is consistent with a Maxwellian definition
of the temperature.
We thus have from Eq.(\ref{mom1bis},\ref{8},\ref{9},\ref{10}) 
\beq
M(1+\ee^2){d\over dt}\vv =\left(p_{[1]}(t)-p_{[2]}(t)\right)
-\left(\lll_{[1]}(V,t)+\lll_{[2]}(V,t)\right)\vv+F_{\rm st}(t),
\label{fst}
\eeq
and 
\bea
F^{[1]\to p}(t)&=&-F^{p\to [1]}(t)={1\over 1+\ee^2}\left[p_{[1]}(t)
-\lambda_{[1]}(V,t)\vv
+F_{\rm st}^{[1]}\right]\label{51}\\
F^{[2]\to p}(t)&=&-F^{p\to [2]}(t)={1\over 1+\ee^2}\left[-p_{[2]}(t)
-\lambda_{[2]}(V,t)\vv
+F_{\rm st}^{[2]}\right]
\eea
where the stochastic forces are
\bea
F_{\rm st}^{[1]}&=&\left[\left\langle V^2\right\rangle_t-\left\langle V\right
\rangle_t^2\right]
 2m n_{[1]}(t)\left\langle v^0\right\rangle^{[1]}_t-
\sum_{k=0}^\infty \left[\left\langle V^{k+3}\right\rangle_t-
\left\langle V\right\rangle_t^{k+3}\right]{4m\over (k+3)!}
n_{[1]}(t)\phi_{[1]}^{\prime(k)}(t)\non
F_{\rm st}^{[2]}&=&-\left[\left\langle V^2\right\rangle_t-\left\langle V\right
\rangle_t^2\right]
 2m n_{[2]}(t)\left\langle v^0\right\rangle^{[2]}_t-
\sum_{k=0}^\infty \left[\left\langle V^{k+3}\right\rangle_t-
\left\langle V\right\rangle_t^{k+3}\right]{4m\over (k+3)!}
n_{[2]}(t)\phi_{[2]}^{\prime(k)}(t).
\eea
and $F_{\rm st}=F_{\rm st}^{[1]}+F_{\rm st}^{[2]}$.

We now compute the heat flux from one side of the piston to the
other side in the microscopic model. This will give us a
microscopic definition of the heat conductivity $\kappa$. We will see
that the microscopical equations are compatible with the
phenomenological thermodynamical equations in which we take into account 
a heat flux characterized by a conductivity $\kappa\ne 0$ 
which comes from the
fluctuations of the microscopical model.
From the first law of thermodynamics 
\beq
{d\over dt}E_{[1]}=P_W^{p\to [1]}+P_Q^{[2]\to [1]}
\eeq
where $P_W^{p\to [1]}$, respectively $P_Q^{[2]\to [1]}$, 
denote the power transmitted 
to the fluid on the left in the form of work, resp. in the form of heat, 
together with the thermodynamic definition of the work power, {\sl i.e.}
\beq
P_W^{p\to [1]}(t)=F^{p\to [1]}(t)V(t)
\eeq
we are led to define, through Eqs.(\ref{51},\ref{16},\ref{17},\ref{serie}),
the microscopic heat power, or heat flux, and the 
microscopic heat conductivity $\kappa$ 
\bea
P_Q^{[2]\to [1]}(t)&=&\left\langle {dE_{[1]}\over dt}\right\rangle
-F^{p\to [1]}(t)\lla V\rra_t\nonumber\\
&=& -{1\over 2}MA_{2,0}^{[1]}+M\sum_{r=0}^\infty \left[A_{1,r}^{[1]}
\left\langle V^r\right\rangle_t\vv-{1\over 2}A_{2,r+1}^{[1]}\left\langle
 V^{r+1}\right\rangle_t
\right]\nonumber\\
&=& M{4\ee^2\over 1+\ee^2}K_0^{[1]}\vv-M{4\ee\over (1+\ee^2)^2}
\left\{\left[(1+\ee^2)\vv^2-(1-{\ee^2\over 2})\vvde\right]K_{-1}^{[1]}
+{1\over 2}K_{1}^{[1]}\right\}\nonumber \\
&& \quad +M\sum_{r=2}^\infty \left[A_{1,r}^{[1]}\left\langle 
V^r\right\rangle_t\vv
-{1\over 2}A_{2,r+1}^{[1]}\left\langle V^{r+1}\right\rangle_t\right]\non
&=&\kappa (T_{[2]}-T_{[1]})
\label{heatflux}\eea
where $A_{2,r}^{[i]}$ are defined by Eq.(\ref{34}) with $K_{1-r}^{[i]}$.

At order ${\cal O}(\ee)$, it reduces to
\bea
P_Q^{[2]\to [1]}&=& 
4\ee MK_{-1}^{[1]}\left[\vvde-\vv^2\right]-2\ee MK_1^{[1]}+{\cal O}(\ee^2)
\nonumber\\
&=&
4{\sqrt{m}\over M}n_{[1]}(t)\left\{M\left[\vvde-\vv^2\right]
\left\langle w\right\rangle_t^{[1]}-{1\over 2}\left\langle
 w^3\right\rangle_t^{[1]}\right\}
+{\cal O}(\ee^2).
\label{pq}
\eea

To compare our definition (\ref{heatflux}) with the heat conductivity
obtained in \cite{lebowitz}, let us consider the case of the 
{\sl infinite} system. In this case the coefficients $K_r^{[i]}$ and the 
densities $n_{[i]}$ are independent on time.
As we shall see in Sec.\ref{stationary}, for any stationary state of the
infinite system we have
\beq
\left\langle V^2\right\rangle_\infty-\left\langle V\right\rangle^2_\infty
={K_1\over 2 K_{-1}}+{\cal O}(\ee)
\eeq
which thus gives for a stationary state
\beq
P_Q^{[2]\to [1]}={2\sqrt{m}\over M}n_{[1]}n_{[2]}
{\lla w^3\rra^{[1]}\lla w\rra^{[2]}
-\lla w^3\rra^{[2]}\lla w\rra^{[1]}\over
\lla w^2\rra^{[1]}\lla w\rra^{[2]}
+\lla w^2\rra^{[2]}\lla w\rra^{[1]}}+{\cal O}(\ee^2).
\eeq
In particular, for a Maxwellian distribution of the fluid particles,
this last equation reads
\beq
P_Q^{[2]\to [1]}={k_B\over M}\sqrt{{8m k_B \over \pi}}n_{[1]}n_{[2]} 
{\sqrt{T_{[1]}T_{[2]}}\over n_{[1]}\sqrt{T_{[1]}}+n_{[2]}\sqrt{T_{[2]}}}
(T_{[2]}-T_{[1]})
+{\cal O}(\ee^2).
\eeq
Therefore the heat conductivity is 
\beq
\kappa={k_B\over M}\sqrt{{8mk_B\over \pi}}n_{[1]}n_{[2]}
{\sqrt{T_{[1]}T_{[2]}}\over n_{[1]}\sqrt{T_{[1]}}+n_{[2]}\sqrt{T_{[2]}}}
+{\cal O}(\ee^2). 
\label{conductivity}
\eeq
Similarly, for the infinite system the friction coefficient 
(\ref{defcoef}) can be written as
\beq
\lll_{[1]}(V)=n_{[1]}\sqrt{{8k_B T_{[1]} m\over \pi }}-n_{[1]}mV
+n_{[1]}m\sqrt{{2m\over \pi k_B T_{[1]}}}{1\over V}\int_0^V dv\, 
{\rm exp}\left(-{mv^2\over 2k_B T_{[1]}}\right)v^2
\label{friction}
\eeq

It is interesting to note that for a stationary state of the {\sl infinite}
system and Maxwellian distribution for the fluids, 
the expression for the
friction coefficient $\lll(V=0)$ Eq.~(\ref{friction}), and for the 
heat conductivity $\kappa$ Eq.~(\ref{conductivity}) are, to lowest order in 
$\ee$,
identical with those first derived by J..L. Lebowitz for a slightly
different model \cite{lebowitz}. In \cite{lebowitz} the piston is
constrained in a state of mechanical equilibrium ({\sl i.e.} 
$\left\langle V^k\right\rangle=0$ for all odd integer $k$) by mean of 
an external potential and the heat flux is just given by $dE_{[1]}/dt$.
In our model there is no external potential and 
in the stationary state the piston has a constant
non-zero average velocity; we thus have to take into account the work power
in the definition of the heat flux Eq.(\ref{heatflux}). 
We should also remark that in 
\cite{lebowitz} the friction coefficient is introduced via the 
Ornstein-Uhlenbeck process describing the motion of the piston, 
while our definition is associated 
with the mechanical equation of motion Eq.(\ref{eqv}) and thus the factor 
$M^{-1}$ difference between $\lll$ in \cite{lebowitz} and our $\lll(V=0)$, 
Eq.(\ref{friction}). Finally, we want to 
stress that in the definition (\ref{defcoef}) 
and (\ref{heatflux}) of $\lll$ and $\kappa$, the 
fluid particles can have any arbitrary distribution and the system is not
necessarily in a stationary state.

To conclude this section we remark that we can use 
the thermodynamic equation 
\beq
{dE_{[1]}\over dt}= T_{[1]}{dS_{[1]}\over dt} -p_{[1]}{dX\over dt}
+\delta M  {dX\over dt}
{d^2X\over dt^2}
\eeq
to define the time evolution of the microscopic entropy by
\bea
T_{[1]}{dS_{[1]}\over dt}&=&
\left\langle {dE_{[1]}\over dt}\right\rangle +p_{[1]}\vv-m \vv{d\over dt}\vv
\nonumber \\
&=& -{M\over 2}\sum_{k=0}^\infty A^{[1]}_{2,r}\left\langle V^r\right\rangle_t
+p_{[1]}\vv-m \vv{d\over dt}\vv\nonumber \\
&=& {2\sqrt{m}\over M}n_{[1]}\left[2\left\langle w\right\rangle_t^{[1]}M\vvde-
\left\langle w^3\right\rangle_t^{[1]}\right]+{\cal O}(\ee^2)
\eea
{\sl i.e.} with (\ref{defcoef}) and (\ref{pq})
\bea
T_{[1]}{dS_{[1]}\over dt}&=& \lll_{[1]}\vvde-2{\sqrt{m}\over M}n_{[1]}(t)
\left\langle w^3\right\rangle_t^{[1]}+{\cal O}(\ee^2)\nonumber \\
&=& \lll_{[1]}\vv^2+{\lll_{[1]}\over M}\left[M(\vvde-\vv^2)-
{\left\langle w^3\right\rangle_t^{[1]}\over 2\left\langle 
w\right\rangle_t^{[1]}} \right]
+{\cal O}(\ee^2)\nonumber\\
&=& \lll_{[1]}\vv^2+ P_Q^{2\to 1},
\eea 
which is consistent with the thermodynamical equation (\ref{eqs1}).
Again, assuming a Maxwellian distribution for 
the fluid particles, we will have 
\beq
T_{[1]}{dS_{[1]}\over dt}=
\lll_{[1]}\vv^2+{\lll_{[1]}\over M}\left[M(\vvde-\vv^2)-k_BT_{[1]}\right]
+{\cal O}(\ee^2).
\label{68}
\eeq

In summary, if initially $\left\langle V^r\right\rangle_{t=0}=0$ for all
$r\geq 1$, then
as long as the contributions
\beq
{\lll_{[i]}\over M}\left[M(\vvde-\vv^2)-k_BT_{[i]}\right], \quad (i=1,2),
\eeq 
as well as the stochastic force $F_{\rm st}$, can be neglected, 
the time evolution Eqs.(\ref{fst},\ref{68}) 
will co\"\i ncide with the thermodynamic evolution of
the adiabatic piston Eq.(\ref{eqv}-\ref{eqs2}) with 
$\kappa=0$. 
This will be the case as long as 
$\left\langle V^r\right\rangle_t\simeq \vv^r$, and $\vv\gg 
k_B\max(T_{[1]},T_{[2]})/M$.
Furthermore, as long as $\vv^2\ll1$ the evolution is described by the adiabats
$S_{[1]}(t)=S_{[1]}$, $S_{[2]}(t)=S_{[2]}$. Therefore during 
this first stage the piston 
evolves according to (\ref{eqv})
towards a state of mechanical equilibrium, {\sl i.e.} $p_{[1]}\simeq p_{[2]}$,
but not thermal equilibrium $(T_{[1]}\neq T_{[2]})$. As soon as 
the pressures are
approximately equal, the above assumptions are no longer valid, and 
as we shall see in Sec.\ref{timeevol} the 
system will evolve toward a state of thermal equilibrium.
Numerical simulations will be presented in Sec.\ref{simul}
to show the time scales on which these evolutions take place.

\section{Stationary non-equilibrium and equilibrium state}\label{stationary}

\subsection{Preliminary remarks}\label{prelimi}

We have seen in Sec.\ref{arbitrary} 
that the time evolution of the piston is
characterized by the equations 
\beq
{d\over dt}\vvs=C_s+\sum_{r=1}^\infty A_{s,r}\vvr,\quad s\geq 1 
\label{evol}
\eeq
where $C_s(t)=A_{s,0}(t)$ and for any $r\geq 0$ 
\beq
A_{s,r}(t)={\ee^{|s-r-1|}\over (1+\ee^2)^s}P_{s,r}(\ee^2)K_{s-r-1}(t).
\eeq
In particular for $s=1$ and $s=2$ we have 
(where $\mxw$ denotes ``for Maxwellian
distributions''), 
\bea
{d\over dt}\lla V\rra_t &=& {1\over M(1+\ee^2)}\left[(p_{[1]}-p_{[2]})
-\lll\lla V\rra_t+m(n_{[1]}-n_{[2]})\lla V^2\rra_t+\cdots\right]
\label{701}\\
{d\over dt}\lla V^2\rra_t &=& {2\over M(1+\ee^2)^2}\left[(1-2\ee^2)
(p_{[1]}-p_{[2]})\lla V\rra_t-\lll\lla V^2\rra_t+2\ee M K_1+\cdots\right]
\label{702}
\eea
with
\bea
p_{[1]}-p_{[2]} &=& M2K_0\mxw k_B(n_{[1]}T_{[1]}-n_{[2]}T_{[2]})\non
\lll &=& M4\ee K_{-1}\mxw \sqrt{m}\sqrt{{8k_B\over \pi}}
\left(n_{[1]}\sqrt{T_{[1]}}+n_{[2]}\sqrt{T_{[2]}}\right)\non
2M\ee K_1 &\mxw & {\sqrt{m}\over M}\sqrt{{8k_B\over \pi}}k_B
\left(n_{[1]}T_{[1]}^{3/2}+n_{[2]}T_{[2]}^{3/2}\right)\non
2K_{-2} &\mxw & n_{[1]}-n_{[2]}
\label{71a}
\eea

To lowest order, {\sl i.e.} in the limit $\ee\to 0$, Eq.(\ref{evol})
together with the definition (\ref{tempdefine}) of the pressure yields
\bea
{d\over dt}\vvs &=& 2sK_0(t)\left\langle V^{s-1}\right\rangle_t\nonumber\\
&=& s{1\over M}[p_{[1]}(t)-p_{[2]}(t)]\lla V^{s-1}\rra_t.
\eea
As discussed in Sec.\ref{micro}, 
for an infinite system with $\ee\ll 1$, the
assumptions
\beq
n_{[i]}(t)=n_{[i]},\quad \phi_{[i]}(v,t)=\phi_{T_{[i]}}(v),\quad(i=1,2)
\eeq
are justified and give $p_{[1]}(t)=p_{[1]}$, $p_{[2]}(t)=p_{[2]}$ constant.

\noi
In this case, for the initial conditions 
\beq
\left\langle V^{s}\right\rangle_{t=0}
=\left(\left\langle V\right\rangle_{t=0}
\right)^s
\eeq
we obtain at lowest order
\bea
\vv &=&\left\langle V\right\rangle_{t=0}+{1\over M}(p_{[1]}-p_{[2]})t
\nonumber\\
\vvs &=&\left(\vv\right)^s
\eea
and thus there is no stationary state in the {\sl infinite} system 
unless $p_{[1]}=p_{[2]}$ in the limit $\ee\to 0$.

On the other hand, for the finite system, if the evolution is sufficiently
slow and $\ee\ll 1$, we may assume that the following homogeneity
conditions hold:
\beq
n_{[1]}(t)={N_{[1]}\over \xx},\quad n_{[2]}(t)= {N_{[2]}\over L-\xx},\quad
\left\langle E_{[i]}\right\rangle_t=N_{[i]}m\left
\langle v^2\right\rangle_t^{[i]}
\equiv {1\over 2}
N_{[i]}k_BT_{[i]}(t),\quad \iii
\label{homogene}
\eeq
where we do not assume Maxwellian distribution of the velocities but 
define the temperatures by the second moments of the fluid velocity 
distribution Eq.(\ref{tempdefine}). 
Under this homogeneity condition, the time evolution for 
$T_{[1]}$ and $T_{[2]}$ are therefore
\beq
{dT_{[i]}(t)\over dt}={2\over N_{[i]}k_B}\left\langle 
{d\over dt}E_{[i]}\right\rangle=
-{M\over N_{[i]} k_B}\sum_{r=0}^\infty A_{2,r}^{[i]}\vvr,\quad\iii.
\label{evolt}\eeq
In the limit $\ee\to 0$ we have from 
Eqs.(\ref{evol},\ref{homogene},\ref{evolt})
\bea
M{d^2\over dt^2}\xx &=& k_B\left[{N_{[1]}\over \xx}T_{[1]} 
-{N_{[2]}\over L-\xx}T_{[2]}\right]
\nonumber\\
{dT_{[1]}\over dt} &=& -{2\over \xx}T_{[1]}{d\over dt}\xx \nonumber\\
{dT_{[2]}\over dt} &=& {2\over L-\xx}T_{[2]}{d\over dt}\xx 
\eea
which yields the ``adiabats''
\beq
T_{[1]}(t)\xx^2=\gamma_{[1]}={\rm cte};\quad T_{[2]}(t)(L-\xx)^2
=\gamma_{[2]}={\rm cte}
\eeq
 and the undamped oscillatory motion
\beq
M{d^2\over dt^2}\xx=k_B\left[{N_{[1]}\gamma_{[1]}\over \xx^3}-
{N_{[2]}\gamma_{[2]}\over (L-\xx)^3}\right].
\eeq

In conclusion, to lowest order in $\ee$, there  is no approach toward an 
equilibrium state in the finite system while 
a stationary state in the infinite system 
can exist only if the initial pressures are 
equal $(p_{[1]}=p_{[2]})$. 
This conclusion is consistent with the previous result,
since for finite $M$ 
the condition $\ee\to 0$ means that $m\to 0$ and thus the
friction coefficient and the heat conductivity are both zero.

\subsection{Stationary state of the infinite system}
\label{6.2}

In this section, we discuss the stationary states of the infinite system.
In this case the densities $n_{[1]}$, $n_{[2]}$, and the coefficients
$K_r$ are constant. Moreover we consider the moments $\lla w^q\rra$
which appear in the definition of $K_r$ to be given by the Maxwellian 
distributions {\sl i.e.}
\bea
&& \lla w^0\rra^{[i]}={1\over 2},\quad 
\lla w^1\rra^{[i]}={1\over 2\pi}\sqrt{2k_BT_{[i]}}, \non
&& \lla w^2\rra^{[i]}={1\over 2}k_BT_{[i]},\quad 
\lla w^3\rra^{[i]}={1\over 2\pi}(2k_BT_{[i]})^{3/2},\quad (i=1,2).
\eea

The stationary states of the infinite system 
are given by the solution of the equations
\beq
0=\sum_{r\geq 0}\ee^{|s-r-1|}P_{s,r}(\ee^2)K_{s-r-1}\lla V^r\rra_\infty.
\label{station}
\eeq

From Eqs.(\ref{701},\ref{702}), it follows immediately that
\bea
&& \lla V\rra_\infty = {p_{[1]}-p_{[2]}\over \lll}
+{m\over \lll}(n_{[1]}-n_{[2]})\lla V^2\rra_\infty+\cdots\label{811}\\
&& \lla V^2\rra_\infty\left[1-(1-2\ee^2){p_{[1]}-p_{[2]}\over \lll}
{m\over \lll}(n_{[1]}-n_{[2]})\right]=(1-2\ee^2)
\left({p_{[1]}-p_{[2]}\over \lll}\right)^2 +2{\ee\over\lll}MK_1+\cdots
\label{812}
\eea 
{\sl i.e.} at lowest order in $\ee$ we have
\bea
\lla V\rra_\infty &=& {p_{[1]}-p_{[2]}\over \lll}+
\left({p_{[1]}-p_{[2]}\over \lll}\right)^2{m\over \lll}{1\over k_B}
\left({p_{[1]}\over T_{[1]}}-{p_{[2]}\over T_{[2]}}\right)+{m\over M}
{p_{[1]}T_{[2]}-p_{[2]}T_{[1]}\over \lll \sqrt{T_{[1]}T_{[2]}}}\label{813}\\
\lla V^2\rra_\infty &=& \left({p_{[1]}-p_{[2]}\over \lll}\right)^2+
{k_B\over M}\sqrt{T_{[1]}T_{[2]}}\label{814}
\eea
which shows that is impossible to have a stationary state with 
$\lla V\rra_\infty=0$, $p_{[1]}=p_{[2]}$, $T_{[1]}\ne T_{[2]}$.

In the following we shall assume that the solution of Eq.(\ref{station}) has an
asymptotic expansion in $\ee$
\beq
\lla V^r\rra_\infty=\sum_{l=0}^\infty \ee^l \lla V^r\rra_\infty^{(l)}.
\label{assumption2}
\eeq
Then in the limit $\ee\to 0$ we must have
\beq
K_0=0,\quad {\sl i.e.}\quad p_{[1]}=p_{[2]}
\label{presseq}
\eeq
which can also be seen explicitly with Eqs.(\ref{811},\ref{812}) since
$\lll={\cal O}(\ee)$.

To take into account the possibility of non-zero pressure difference for
the infinite system, and to have results valid for finite systems where
the equilibrium densities might depend on $\ee$, we now consider that
the densities $n_{[1]}$, $n_{[2]}$ are function of $\ee$
\beq
n_{[i]}(\ee)=\sum_{l=0}^\infty \ee^l n_{[i]}^{(l)}\quad (i=1,2)
\eeq
which then implies that the coefficients $K_r$ are also functions of $\ee$
\beq
K_r(\ee)=\sum_{l=0}^\infty \ee^l K_r^{(l)}
\eeq
The condition (\ref{presseq}) for a stationary solution is now
\beq
K_0^{(0)}=\lim_{\ee\to 0} K_0(\ee)=0
\eeq
{\sl i.e.}
\beq
n_{[1]}^{(0)}T_{[1]}-n_{[2]}^{(0)}T_{[2]}=0
\eeq
or
\beq
\lim_{\ee\to 0}[p_{[1]}(\ee)-p_{[2]}(\ee)]=p_{[1]}^{(0)}-p_{[2]}^{(0)}=0.
\eeq
Therefore Eq.(\ref{station}) can be simplified by $\ee$ to give
\bea
0&=& P_{s,s-1}\left({K_0\over \ee}\right)\lla V^{s-1}\rra_\infty+\nonumber\\
& &\quad + \sum_{r=0}^{s-2}\ee^r \left[P_{s,s-2-r}K_{1+r}
\lla V^{s-2-r}\rra_\infty
+P_{s,s+r}K_{-1-r}\lla V^{s+r}\rra_\infty\right]\nonumber\\
& &\quad + \sum_{r=0}^{\infty}\ee^{s-1+r}P_{s,2s-1+r}K_{-s-r}
\lla V^{2s-1+r}\rra_\infty.
\label{station2}
\eea
We then solve (\ref{station2}) at successive order in  $\ee$.

At order $\ee^0$, we have 
\bea
0&=& 2K_0^{(1)}-4K_{-1}^{(0)}\lla V\rra_\infty^{(0)}\quad(s=1)\nonumber\\
0&=& 4K_0^{(1)}\lla V\rra_\infty^{(0)}+4K_1^{(0)}-8K_{-1}^{(0)}
\lla V^{2}\rra_\infty^{(0)}\quad(s=2)\non
0&=& 2sK_0^{(1)}\lla V^{s-1}\rra_\infty^{(0)}+2s(s-1)K_1^{(0)}
\lla V^{s-2}\rra_\infty^{(0)}
-4sK_{-1}^{(0)}\lla V^{s}\rra_\infty^{(0)}\quad(s\geq 3)
\eea
which yields 
\bea
\lla V\rra_\infty^{(0)} &=& {K_0^{(1)}\over 2 K_{-1}^{(0)}}=V_0
={1\over 2\sqrt{M}}
{n_{[1]}^{(1)}\lla w^2 \rra^{[1]}-n_{[2]}^{(1)}\lla w^2 \rra^{[2]}\over 
n_{[1]}^{(0)}\lla w \rra^{[1]}+n_{[2]}^{(0)}\lla w \rra^{[2]}}\nonumber\\
&=& \sqrt{{\pi k_B \over 8M}T_{[1]}T_{[2]}}\lim_{\ee\to 0}{1\over \ee}
{p_{[1]}(\ee)-p_{[2]}(\ee)\over p_{[1]}(\ee)\sqrt{T_{[2]}}
+p_{[2]}(\ee)\sqrt{T_{[1]}}},
\label{91}
\eea
\bea
\lla V^2\rra_\infty^{(0)}-\left(\lla V\rra_\infty^{(0)}\right)^2 &=& 
 {K_1^{(0)}\over 2 K_{-1}^{(0)}}=\Delta={1\over 2M}
\lim_{\ee\to 0}{n_{[1]}(\ee)\lla w^3 \rra^{[1]}
+n_{[2]}(\ee)\lla w^3 \rra^{[2]}\over 
n_{[1]}(\ee)\lla w \rra^{[1]}+n_{[2]}(\ee)\lla w \rra^{[2]}}\non
&=& {k_B\over M}\sqrt{T_{[1]}T_{[2]}}
\lim_{\ee\to 0}
{p_{[1]}(\ee)\sqrt{T_{[1]}}+p_{[2]}(\ee)\sqrt{T_{[2]}}
\over p_{[1]}(\ee)\sqrt{T_{[2]}}
+p_{[2]}(\ee)\sqrt{T_{[1]}}},
\label{v2}
\eea
and
\beq
\lla V^s\rra_\infty^{(0)}
=\sum_{k=0}^{[s/2]}{s!\over 2^k k!(s-2k)!}V_0^{s-2k}s^k.
\eeq
Therefore at order $\ee^0$, the distribution function for the velocity of the
piston is 
\beq
\Phi^{(0)}(V)={1\over \sqrt{2\pi \Delta}}
\exp\left[-{(V-V_0)^2\over 2\Delta}\right].
\eeq
Let us remark that we recover Eqs.(\ref{813},\ref{814}) at order
zero. 

At order $\ee^1$, introducing the notation 
\beq
k_r^{(l)}={K_r^{(l)}\over 2K_{-1}^{(0)}};\quad k_0^{(0)}=0;
\quad k_0^{(1)}=V_0;\quad k_1^{(0)}=\Delta 
\eeq
Eq.(\ref{station2}) yields
\bea
\lla V\rra_\infty^{(1)} &=& -2k_{-1}^{(1)}V_0+k_{0}^{(2)}+k_{-2}^{(0)}
(\Delta+V_0^2)\label{102}\\
\lla V^2\rra_\infty^{(1)} &=& V_0\lla V\rra_\infty^{(1)}-2k_{-1}^{(1)}
(\Delta+V_0^2)
+k_{0}^{(2)}V_0+k_1^{(1)}+k_{-2}^{(0)}V_0(3\Delta+V_0^2)\\
\lla V^s\rra_\infty^{(1)} &=& V_0\lla V^{s-1}\rra_\infty^{(1)}+(s-1)\Delta\lla 
V^{s-2}\rra_\infty^{(1)}-2k_{-1}^{(1)}\lla V^{s}\rra_\infty^{(0)}
+k_{0}^{(2)}\lla V^{s-1}\rra_\infty^{(0)}\nonumber\\
&& \quad+(s-1)k_{1}^{(1)}\lla V^{s-2}\rra_\infty^{(0)}
+{2\over 3}(s-1)(s-2)k_{2}^{(0)}\lla V^{s-3}\rra_\infty^{(0)}
+k_{-2}^{(0)}\lla V^{s+1}\rra_\infty^{(0)}.
\eea
As one can verify the distribution function at order $\ee^1$ is 
\bea
\Phi_\ee(V) &=&{1\over \sqrt{2\pi\Delta}}\exp\left[-{1\over 2\Delta}
\left(V-\sqrt{{\pi k_B \over 8M}T_{[1]}T_{[2]}}
{p_{[1]}(\ee)-p_{[2]}(\ee)\over p_{[1]}(\ee)\sqrt{T_{[2]}}
-p_{[2]}(\ee)\sqrt{T_{[1]}}}\right)^2\right]
\nonumber\\
&&\quad\times \left\{ 1+\ee\left[a_1(V-V_0)+a_2(V^2-\Delta-V_0^2)
+a_3(V^3-3\Delta V_0 -V_0^3)\right]\right\}
\label{101}
\eea
where $V_0$, $\Delta$ are given by Eqs.(\ref{91},\ref{v2}) and 
\bea
a_1\Delta^3 &=&k_{-1}^{(1)}2V_0\Delta^2-k_{1}^{(1)}V_0\Delta
-{2\over 3} k_{2}^{(0)}(\Delta-V_0^2)\\
a_2\Delta^3 &=& -k_{-1}^{(1)}\Delta^2+{1\over 2}k_{1}^{(1)}\Delta
-{2\over 3} k_{2}^{(0)}V_0\\
3a_3\Delta^3 &=&k_{-2}^{(0)}\Delta^2+{2\over 3} k_{2}^{(0)}.
\eea
Let us remark that for $p_{[1]}=p_{[2]}$ then $V_0=0$ and 
we recover the results of Gruber and
Piasecki \cite{gruber1}. 
One can then successively obtain higher order terms for the velocity
moments.

\noi In conclusion, for the infinite system we observe a stationary state 
with non-zero average 
velocity $\lla V\rra_\infty$, 
which depends on $n_{[1]}$, $n_{[2]}$, $T_{[1]}$,
$T_{[2]}$, 
given at order $\ee$ by Eqs.(\ref{813},\ref{102}).

It is interesting to note that this result present some analogy and
differences with those obtained by J.L. Lebowitz in \cite{lebowitz}, where
he considered an adiabatic piston constrained around the origin by 
an external force: The coefficient $\Delta$ Eq.(\ref{v2}) is exactly
the same as in \cite{lebowitz}; however for 
$p_{[1]}-p_{[2]}={\cal O}(\ee)\ne 0$ we have a drift
$V_0$, while in \cite{lebowitz} the external force give 
$\lla V^{2k+1}\rra_\infty^{(0)}=0$ ($k\geq 0$).

To conclude the discussion on the stationary states of the infinite system, 
we investigate under what condition there exists a stationary 
state with zero average velocity {\sl i.e.} $\lla V\rra_\infty=0$.

From Eq.(\ref{811}) it follows immediately that
\beq
p_{[1]}-p_{[2]}=-m(n_{[1]}-n_{[2]})\lla V^2\rra_\infty+{\cal O}(\ee^3).
\eeq
However one has to check that a stationary solution of Eq.(\ref{station}) does 
exist with $\lla V\rra_\infty=0$ at all order in $\ee$. 

From Eqs.(\ref{91},\ref{102}), $\lla V\rra_\infty=0$ up to order $\ee$ iff
\beq
K_0^{(0)} = K_0^{(1)}=0,\quad{\sl i.e.}\quad 
\lim_{\ee\to 0}(p_{[1]}(\ee)-p_{[2]}(\ee))/\ee=0
\label{equal1}
\eeq
and
\beq
K_0^{(2)} + K_{-2}^{(0)}\Delta =0.
\eeq
Expressed differently, $\lla V\rra_\infty=0$ up to order $\ee$ iff
\bea
&& n_{[1]}^{(0)}T_{[1]}-n_{[2]}^{(0)}T_{[2]}=0\nonumber\\
&& n_{[1]}^{(1)}T_{[1]}-n_{[2]}^{(1)}T_{[2]}=0\nonumber\\
&& n_{[1]}^{(2)}T_{[1]}-n_{[2]}^{(2)}T_{[2]}=-{M\over k_B}\Delta
\left[n_{[1]}^{(0)}-n_{[2]}^{(0)}\right].
\label{equal2}
\eea
This last equation is equivalent to
\beq
p_{[1]}-p_{[2]}=-m\left[n_{[1]}^{(0)}
-n_{[2]}^{(0)}\right]\lla V^2\rra_\infty^{(0)}
\eeq
and, in this case
\bea
\lla V^2\rra_\infty^{(0)} &=& \Delta={k_B\over M}\sqrt{T_{[1]}T_{[2]}}\\
\lla V^2\rra_\infty^{(1)} &=& k_1^{(1)}-2k_{-1}^{(1)}\Delta.
\eea

At order $\ee^2$, we have (taking $s=1$)
\beq
\lla V\rra_\infty^{(2)}=k_0^{(3)}-2k_{-1}^{(1)}
\lla V\rra_\infty^{(1)}-2k_{-1}^{(2)}V_0
+k_{-2}^{(1)}\lla V^2\rra_\infty^{(0)}+k_{-2}^{(0)}\lla V^2\rra_\infty^{(1)}
-{1\over 3}k_{-3}^{(0)}\lla V^3\rra_\infty^{(0)}.
\eeq
Therefore $\lla V\rra_\infty=0$ up to order $\ee^2$ iff
\beq
K_0^{(3)}+K_{-2}^{(1)}\Delta +K_{-2}^{(0)}(k_{1}^{(1)}-2k_{-1}^{(1)}\Delta)=0
\eeq

Finally at order $\ee^q$, we have (taking $s=1$)
\beq
0=2[K_0]^{(q+1)}-4[K_{-1}\lla V\rra_\infty]^{(q)}
+2[K_{-2}\lla V^2\rra_\infty]^{(q-1)}
-\sum_{r\geq 3}{4\over r!}[K_{-r}\lla V_r\rra_\infty]^{(q+1-r)}.
\eeq
Therefore $\lla V\rra_\infty=0$ up to order $\ee^q$ iff
all $K_0^{(l)}$ are uniquely determined by the functions obtained 
at previous order.

In conclusion, under the assumption Eq.(\ref{assumption2}) 
and given the  density
$n_{[2]}$, and the distributions $\phi_{[1]}(v)$,  $\phi_{[2]}(v)$, with
$T_{[1]}\ne T_{[2]}$, then the average velocity of the piston in the 
stationary state is
non-zero for all densities $n_{[1]}$ 
except for one special value $n_{[1]}=n_{[1]}(\ee)$ 
for which $\lla V\rra_\infty=0$.
For this special solution, one has 
\bea
n_{[1]}^{(0)}\lla w^2\rra^{[1]} &=& n_{[2]}\lla w^2\rra^{[2]}\\
n_{[1]}^{(1)} &=& 0\\
n_{[1]}^{(2)}\lla w^2\rra^{[1]} &=& -[n_{[1]}^{(0)}\lla w^0\rra^{[1]}-
n_{[2]}\lla w^0\rra^{[2]}]\Delta\\
{1\over M}n_{[1]}^{(q)}\lla w^2\rra^{[1]} &=& 
-[K_{-2}\lla V^2\rra_\infty]^{(q-2)}
+\sum_{r\geq 3}{2\over r!}[K_{-r}\lla V_r\rra_\infty]^{(q-r)}\quad (q\geq 3).
\eea
In this state we have (see Sec.\ref{pheno})
\beq
P_Q^{2\to 1}=-{1\over \sqrt{M}}2\ee n_{[2]} \lla w^2\rra^{[2]}
{\lla w^3\rra^{[1]}\lla w\rra^{[2]}
-\lla w^3\rra^{[2]}\lla w\rra^{[1]}\over
\lla w^2\rra^{[1]}\lla w\rra^{[2]}
+\lla w^2\rra^{[2]}\lla w\rra^{[1]}}
+{\cal O}(\ee^2)
\eeq
and thus there is a constant heat flux for any distribution $\phi_{[1]}(v)$, 
$\phi_{[2]}(v)$ such that $\lla w^3\rra^{[1]}\lla w\rra^{[2]}\ne
\lla w^3\rra^{[2]}\lla w\rra^{[1]}$.
For Maxwellian distribution, this means that there is a constant flux 
for any $T_{[1]}\ne T_{[2]}$.

\noi Therefore there exists a special non-equilibrium stationary state with 
zero average velocity of the piston, which is however distinct 
from the one given in \cite{lebowitz} since in our case 
$\lla V^{2k+1}\rra_\infty$ is not zero (for $k\geq 1$).

\subsection{Equilibrium state of the finite system}
\label{equilfinite}

Any equilibrium state of the finite  system is necessarily a stationary 
state with $\lla V\rra_\infty=0$ and $P_Q^{[2]\to [1]}=0$ where, for
$\lla V\rra_\infty =0$ (see Eq.(\ref{heatflux})),
\bea
P_Q^{[2]\to [1]}=-{2M\over (1+\ee^2)^2}\left[\ee K_1^{[1]}(\infty) \right.
&-& 
2\ee(1-\ee^2/2)
K_{-1}^{[1]}(\infty)\lla V^2\rra_\infty+\ee^2K_{-2}^{[1]}(\infty)
\lla V^3\rra_\infty
\nonumber\\
& & \left.
-\sum_{r\geq 4}\ee^{r-1}{2\over r!}\left(r+\ee^2(r-3)\right)K^{[1]}_{-r+1}
(\infty)
\lla V^r\rra_\infty\right].
\label{heatstation}
\eea
Taking into account (\ref{v2}), Eq.(\ref{heatstation}) implies at order $\ee$
\beq
\lla w^3\rra_\infty^{[1]}\lla w\rra_\infty^{[2]}=
\lla w^3\rra_\infty^{[2]}\lla w\rra_\infty^{[1]}.
\eeq
Assuming Maxwellian distribution for the fluid particles, 
it implies $T_{[1]}(\infty)=T_{[2]}(\infty)$ at order $\ee$
and in this case we have
\beq
\phi_{[1]}(v)=\phi_{[2]}(-v).
\label{equaldistr}
\eeq
We conjecture that for any equilibrium state 
the relation (\ref{equaldistr}) must be satisfied,
which then implies $T_{[1]}(\infty)=T_{[2]}(\infty)$.

Assuming this conjecture (\ref{equaldistr}) to hold, Eq.(\ref{equal1}) yields 
first 
\bea
&& K_0^{(l)}(\infty)=0\quad {\rm for}\quad l=0,1\\
{\sl i.e.} \quad && n_{[1]}^{(l)}(\infty)= n_{[2]}^{(l)}(\infty)
\quad {\rm for}\quad l=0,1
\label{n1n2}
\eea
Then, with Eq.(\ref{equal2}), Eq.(\ref{n1n2}) is also valid for $l=3$, thus
\beq
K_r^{(l)}(\infty)=0\quad {\rm for\ all\ }r{\rm\  even},\quad l=0,1,2
\eeq
and with (\ref{101})
\beq
\lla V^{2k+1}\rra_\infty^{(1)}=0 \quad{\rm for\ all\ }k\geq 0.
\eeq 
Iterating the argument at successive order we arrive at the following
conclusion. Under the assumption (\ref{assumption2}) and the conjecture
(\ref{equaldistr}), any equilibrium state of the finite system must satisfy 
$n_{[1]}(\infty)=n_{[2]}(\infty)$, 
in other words $p_{[1]}(\infty)=p_{[2]}(\infty)$ 
and $T_{[1]}(\infty)=T_{[2]}(\infty)$.

Conversely, if $\phi_{[1]}(v)=\phi_{[2]}(-v)$, thus $T_{[1]}=T_{[2]}$, 
and $n_{[1]}(\infty)
=n_{[2]}(\infty)$ then the solution of
the stationary equation at order $\ee$ is 
\bea
\lla V^{2k+1}\rra_\infty&=& 0\nonumber\\
\lla V^{2k}\rra_\infty &=& (2k-1)!!\Delta^k,\quad \Delta={K_1(\infty)
\over 2K_{-1}(\infty)}.
\eea
Therefore at order $\ee$ the velocity distribution of the piston 
is necessarily Maxwellian with temperature $T=M\Delta/k_B$, 
{\sl i.e.}
\beq
\Phi_\ee(V) =\sqrt{{1\over 2\pi\Delta}}\exp\left[-{V^2\over 2\Delta}\right].
\eeq

In conclusion the adiabatic piston is stricto senso a conductor since the 
only equilibrium states are those for which $T_{[1]}(\infty)
=T_{[2]}(\infty)$. 
However the important 
question to be considered in the next section 
concerns the time scale necessary to reach this true equilibrium 
state.

\section{Time evolution}\label{timeevol}

In this section we give a qualitative (non rigorous) discussion of the time 
evolution for the infinite and the finite systems.

\subsection{Infinite systems}

As discussed in Sec.\ref{micro}, for infinite systems and $\ee\ll 1$ 
the assumptions $n_{[i]}(t)=n_{[i]}$, $\phi_{[i]}(v,t)
=\phi_{T_{[i]}}(v)$, $i=1,2$ are justified. In this case the functions 
$K_r$ are constant, and we are led to solve the equation (\ref{serie})
\beq
{d\over dt}{\cal V}=\ee\left[{\cal C}+{\cal AV}\right],
\quad {\cal V}(t=0)=0,
\eeq
where 
\bea
{\cal V}&=&\left\{\lla V^s\rra_t;\ s\geq 1\right\}\non
{\cal C}_s&=&{1\over \ee}A_{s,0}=2^s\ee^{s-1}M^{-{s+1\over 2}}\left[
n_{[1]}\lla w^{s+1}\rra^{[1]}+n_{[2]}\lla w^{s+1}\rra^{[2]}\right]\non
{\cal A}_{s,r}&=&{\ee^{|s-r-1|}\over (1+\ee^2)^s}{1\over \ee}P_{s,r}(\ee^2)
K_{s-r-1}
\eea
{\sl i.e.} with $t'=\ee t$
\beq
{d\over dt'}{\cal V}={\cal C}+{\cal AV}.
\eeq
In the limit $\ee\to 0$, with the condition that 
\beq
\lim_{\ee\to 0}{1\over \ee}K_0=\lim_{\ee\to 0}{1\over \ee}\left[p_{[1]}(\ee)-
p_{[2]}(\ee)\right]
\eeq
exists and is finite, the matrix ${\cal A}_{s,r}$ is non-zero
only for $r=s-2,s-1,s$, and its eigenvalues $\alpha_n$ are all negative
given by
\beq
\alpha_s=-s4K_{-1}=-s{\lll\over \ee M},\quad s\geq 1
\eeq
where the friction coefficient $\lll=\lll(V=0)$, given by (\ref{friction}), is
of order $\ee$. 

\noi
We thus obtain at lowest order in $\ee$ 
\bea
\lla V\rra^{(0)}_t &=& V_0\left(1-{\rm e}^{-{\lll\over M}t}\right)\non
\lla V^2\rra^{(0)}_t &=& V_0^2\left(1-{\rm e}^{-{\lll\over M}t}\right)^2
+\Delta\left(1-{\rm e}^{-2{\lll\over M}t}\right) 
\eea
with $V_0$ and $\Delta$ given by Eqs.(\ref{91},\ref{v2}) or 
(\ref{91},\ref{v2}).

\noi
At order $\ee^2$ we find for
the case $p_{[1]}=p_{[2]}$ 
\bea
\lla V\rra_t &=& \lla V\rra_\infty
\left(1-{\rm e}^{-{\lll\over M(1+\ee^2)}t}\right)^2\label{139}\\
\lla V^2\rra_t &=&\lla V^2\rra_\infty
\left[1-{\rm e}^{-2{\lll\over M(1+\ee^2)}t}+2\ee^2
\left({\rm e}^{-{\lll\over M(1+\ee^2)}t}-{\rm e}^{-2{\lll\over M(1+\ee^2)}t}
\right)\right]
\eea
with $\lla V\rra_\infty$ and $\lla V^2\rra_\infty$ given by Eqs.(\ref{813},
\ref{814}) with $p_{[1]}=p_{[2]}$.

Similarly for $p_{[1]}\ne p_{[2]}$ we will have another linear combination of 
$\exp(-t/\tau_a)$, $\exp(-2t/\tau_a)$ with 
\beq
\tau_a={M+m\over \lll}.
\label{7.1.1}
\eeq
We notice that taking realistic numbers one finds $\tau_a$ of the order 
$10^{-1}-10^{-2}$ which means that the stationary state is reached 
very rapidly.

\subsection{Finite systems}
\label{7.2}

For finite systems the equations for the time evolution of 
$\lla V\rra_\infty$ and $\lla V^2\rra_\infty$ are given at order 
$\ee^2$ by Eqs.(\ref{701},\ref{702}) where the densities $n_{[i]}$ and
the coefficients $K_r^{[i]}$ are now functions of time.
Since we do not consider the full many-body problem describing the fluids,
we take care of this unknown time dependence by assuming that the 
fluids satisfy the homogeneity condition (\ref{homogene}). Moreover for
$K_r^{[i]}$, $r=-2,-1,1$, we take the expressions (\ref{71a}) obtained 
with the Maxwellian distributions. From Eq.(\ref{evolt}) we obtain 
at order $\ee$
\bea
{d\over dt}T_{[1]} &=& -{2T_{[1]}\over \lla X\rra_t}\left[
\lla V\rra_t-\sqrt{m}\sqrt{{8\over \pi k_BT_{[1]}}}\lla V^2\rra_t
+{\sqrt{m}\over M}\sqrt{{8k_B T_{[1]}\over \pi}}\right]\label{7.2.1}\\
{d\over dt}T_{[2]} &=& -{2T_{[2]}\over L-\lla X\rra_t}\left[
-\lla V\rra_t-\sqrt{m}\sqrt{{8\over \pi k_BT_{[2]}}}\lla V^2\rra_t
+{\sqrt{m}\over M}\sqrt{{8k_B T_{[2]}\over \pi}}\right]\label{7.2.2}
\eea
At order $\ee$ we thus have a system of five O.D.E for the unknown
$\lla X\rra_t$, $\lla V\rra_t$, $\lla V^2\rra_t$, $T_{[1]}$ and $T_{[2]}$.
At this order we conclude that the system will evolve toward the unique
equilibrium state
\bea
T_{[1]}(\infty)=T_{[2]}(\infty) &=& {N_{[1]}T_{[1]}+N_{[2]}T_{[2]}\over
N_{[1]}+N_{[2]}}\\
\lla X\rra_\infty &=& L{N_{[1]}\over N_{[1]}+N_{[2]}}\\
\lla V\rra_\infty &=& 0\\
\lla V^2\rra_\infty &=& {k_B\over M}{N_{[1]}T_{[1]}+N_{[2]}T_{[2]}\over
N_{[1]}+N_{[2]}}
\eea
which is identical to the one obtained in thermodynamics for the conducting
piston ($\kappa\ne 0$).

Let us then analyze qualitatively how the evolution toward this final
equilibrium state takes place.

Given that at $t=0$, we have $X=X_0$, $V=0$, $T_{[1]}(0)=T_{[1]}$,
$T_{[2]}(0)=T_{[2]}$, and $p_{[1]}(0)\ne p_{[2]}(0)$, then as long
as the velocity $\lla V\rra_t$ remains small Eqs.(\ref{7.2.1},\ref{7.2.2})
yield the adiabats
\bea
T_{[1]}(t)\lla X\rra_t^2 &=& T_{[1]}X_0^2\label{7.2.7}\\
T_{[2]}(t)\left(L-\lla X\rra_t\right)^2 &=& T_{[2]}\left(L-X_0\right)^2
\label{7.2.8}
\eea
and the time evolution for the piston is 
\beq
{d\over dt}\lla V\rra_t={k_B\over M}\left[
N_{[1]}T_{[1]}{X_0^2\over \lla X\rra_t^3}
-N_{[2]}T_{[2]}{\left(L-X_0\right)^2\over \left(L-\lla X\rra_t\right)^3}
\right]
-{\lll(t)\over M}\lla V\rra_t
\label{149}
\eeq
with $\lll(t)$ given by Eq.(\ref{71a}), together with Eqs.(\ref{7.2.7},
\ref{7.2.8}).

The piston evolves thus adiabatically until a time $t_a$, of the order
$\tau_a$ Eq.(\ref{7.1.1}), where the pressures become equal, {\sl i.e.}
\beq
{N_{[1]}T_{[1]}(t_a)\over \lla X\rra_{t_a}}=
{N_{[2]}T_{[2]}(t_a)\over L-\lla X\rra_{t_a}}
\eeq
which yields
\bea
\lla X\rra_{t_a} &=&{L\over 1+\left[{N_{[2]}T_{[2]}\over N_{[1]}T_{[1]}}
\left({L-X_0\over X_0}\right)^2\right]^{1/3}}\label{151}\\
T_{[1]}(t_a)&=&T_{[1]}\left({X_0\over \lla X\rra_{t_a}}\right)^2\\
T_{[2]}(t_a)&=&T_{[2]}\left({L-X_0\over L-\lla X\rra_{t_a}}\right)^2\\
p(t_a)&=&k_BN_{[1]}T_{[1]}{X_0^2\over \lla X\rra_{t_a}^3}.\label{154}
\eea

In the next stage, as can be seen from the numerical simulations of
Sec.\ref{simul}, the pressures will remain approximatively constant
and equal, {\sl i.e.}
\bea
p_{[1]}(t) &=& k_B{N_{[1]}T_{[1]}(t)\over \lla X\rra_t}
\simeq {\rm cte}\label{7.2.15}\\ 
p_{[2]}(t) &=& k_B{N_{[2]}T_{[2]}(t)\over L-\lla X\rra_t}\simeq {\rm cte}
\label{7.2.16}\eea
and 
\beq
N_{[1]}T_{[1]}(t)\left[L- \lla X\rra_t\right]=N_{[2]}T_{[2]}(t)\lla X\rra_t
\label{7.2.17}
\eeq
From Eqs.(\ref{7.2.15},\ref{7.2.16}) we have
\bea
{1\over T_{[1]}}{d\over dt}T_{[1]} &=& {\lla V\rra_t\over \lla X\rra_t}\\
{1\over T_{[2]}}{d\over dt}T_{[2]} &=& {\lla V\rra_t\over L-\lla X\rra_t}
\eea
and using Eqs.(\ref{7.2.1},\ref{7.2.2}) we conclude that 
\bea
M\lla V^2\rra_t &=& k_B \sqrt{T_{[1]}(t)T_{[2]}}(t)\label{7.2.20}\\
\lla V\rra_t &=& {2\over 3} {\sqrt{m}\over M}\sqrt{{8k_B\over \pi}}\left[
\sqrt{T_{[2]}(t)}-\sqrt{T_{[1]}(t)} \right].
\label{7.2.21}
\eea
In other words the temperatures $T_{[1]}(t)$ and $T_{[2]}(t)$ 
evolves, but at all time $t$ we have
\bea
\lla V^2\rra_t &=&\lla V^2\rra_\infty\\
\lla V\rra_t &=&{16\over 3\pi}\lla V\rra_\infty
\eea
where $\lla V\rra_\infty$ and $\lla V^2\rra_\infty$ are the stationary
values for the infinite system with temperatures $T_{[i]}=T_{[i]}(t)$, 
$i=1,2$.

\noi
Introducing the expressions (\ref{7.2.20}) and (\ref{7.2.21}) in 
(\ref{7.2.1}), (\ref{7.2.2}) yields, with $p_{[1]}=p_{[2]}=p$
\bea
{d\over dt}T_{[1]}&=&-{2\over 3}{p\over N_{[1]}k_B}{\sqrt{m}\over M}
\sqrt{{8k_B\over \pi}}\left[\sqrt{T_{[1]}}-\sqrt{T_{[2]}}\right]
\label{7.2.22}\\
{d\over dt}T_{[2]}&=&{2\over 3}{p\over N_{[2]}k_B}{\sqrt{m}\over M}
\sqrt{{8k_B\over \pi}}\left[\sqrt{T_{[1]}}-\sqrt{T_{[2]}}\right]
\eea
and thus
\beq
{d\over dt}(T_{[2]}-T_{[1]})=-{2\over 3}\left({1\over N_{[1]}k_B}+
{1\over N_{[2]}k_B}\right)
{\sqrt{m}\over M}
\sqrt{{8k_B\over \pi}}
{n_{[1]}k_B T_{[1]}\over \sqrt{T_{[1]}}+\sqrt{T_{[2]}}}
\left(T_{[2]}-T_{[1]}\right).
\eeq
Let us note that for $p_{[1]}=p_{[2]}$, it follows from 
Eqs.(\ref{conductivity},\ref{friction}) that 
\bea
\kappa &=& {\sqrt{m}\over M}
\sqrt{{8k_B\over \pi}}
{n_{[1]}k_B T_{[1]}\over \sqrt{T_{[1]}}+\sqrt{T_{[2]}}}\\
\lll_{[1]}+\lll_{[2]}&=&\sqrt{m}\sqrt{{8k_B\over \pi}}
n_{[1]}T_{[1]}{\sqrt{T_{[1]}}+\sqrt{T_{[2]}}\over \sqrt{T_{[1]}T_{[2]}}}\\
{\sl i.e.}\quad \kappa &=& {\lll\over M}k_B(\theta+\theta^{-1})^2,\quad
\theta=\left({T_{[1]}\over T_{[2]}}\right)^{1/4}.
\eea
We thus have
\beq
{d\over dt}(T_{[2]}-T_{[1]})=-{2\over 3}
\left({1\over N_{[1]}k_B}+
{1\over N_{[2]}k_B}\right)\kappa\left(T_{[2]}-T_{[1]}\right)
\label{170}
\eeq
from which follows that the system will reach equilibrium with a 
relaxation time $\tau_{\rm e}$ of the order 
\beq
\tau_{\rm e}\simeq \kappa^{-1}\simeq {1\over k_B}{M\over \lll}
\simeq {1\over k_B} \tau_a.
\eeq
Taking realistic numbers we obtain a relaxation time which is several
time the age of the universe. 

The coefficient $2\left({1\over N_{[1]}k_B}+{1\over N_{[2]}k_B}\right)$
is clearly the inverse of the specific heat. To understand the 
origin of the factor $1/3$ we must remember that the piston is moving.
Using Eqs.(\ref{7.2.21},\ref{7.2.22}) yields
\beq
P_Q^{[2]\to [1]}={1\over 2}N_{[1]}k_B {dT_{[1]}\over dt}
+p_{[1]}\lla V\rra_t=\kappa \left(T_{[2]}-T_{[1]}\right)=-P_Q^{[1]\to [2]}
\eeq
as it should.

Let us also note that the equality of pressure (\ref{7.2.17}), together
with the conservation of energy
\beq
N_{[1]}T_{[1]}(t)+N_{[2]}T_{[2]}(t)+{M\over k_B}\lla V^2\rra_t=
N_{[1]}T_{[1]}+N_{[2]}T_{[2]},
\eeq
which is exactly satisfied by 
Eqs.(\ref{701},\ref{702},\ref{7.2.1},\ref{7.2.2}) if $p_{[1]}=p_{[2]}$,
gives the relation between position and temperatures 
(\ref{7.2.15},\ref{7.2.16})
\bea
LN_{[1]}T_{[1]}(t) &=& \left(N_{[1]}T_{[1]}+N_{[2]}T_{[2]}
-{M\over k_B}\lla V^2\rra_t\right)\lla X\rra_t\non
&\simeq &\left(N_{[1]}T_{[1]}+N_{[2]}T_{[2]}\right)\lla X\rra_t\\
LN_{[2]}T_{[2]}(t) 
&\simeq &\left(N_{[1]}T_{[1]}+N_{[2]}T_{[2]}\right)\left(L-\lla X\rra_t
\right)
\eea
and thus from Eq.(\ref{7.2.21}), or Eq.(\ref{7.2.22}),
\beq
{d\over dt}\lla X\rra_t={2\over 3}{\sqrt{m}\over M}
\sqrt{{8\over \pi}}\sqrt{{k_B \left(N_{[1]}T_{[1]}+N_{[2]}T_{[2]}\right)
\over L}}\left(\sqrt{{L-\lla X\rra_t\over N_{[2]}}}-
\sqrt{{\lla X\rra_t\over N_{[1]}}}\right).
\label{7.2.32}
\eeq

Finally if we compare ${d\over dt}\lla V\rra_t$ obtained 
from Eq.(\ref{7.2.32}) with Eq.(\ref{701}), we arrive at the conclusion
that
\beq
p_{[2]}-p_{[1]}={\cal O}\left(\ee^2(T_{[2]}-T_{[1]})\right)
\eeq
which justifies our starting point for the second stage.

Let us also remark that for some time interval 
between the adiabatic evolution Eq.(\ref{149}), and
the approach toward equilibrium with $p_{[1]}\simeq p_{[2]}$ (\ref{170}), 
there will be an intermediate stage for which $P_Q^{[1]\to[2]}+P_Q^{[2]\to[1]}$
is not zero. During this intermediate evolution the stochastic motion of the
piston will have to be introduced in the thermodynamical equations not only
with a conductivity coefficient $\kappa\ne 0$, but also by means of an 
internal energy $U_p$ and and entropy $S_{p}$ of the piston with
$U_p=U_p(S_p)$.

\section{Numerical simulations}\label{simul}

In order to verify the assumptions on which our previous analysis is based, 
we made numerical simulations for the finite as well as for 
the infinite system in one dimension taking $k_B=1$. The initial state of the 
fluids particles of mass $m$ is
given by Maxwellian distributions of velocities according to Eq.(\ref{1}). 
Then the particles and the piston (a particle of mass $M$ with initial
coordinate $(X_0,V_0)$) interact only through elastic collisions.

\begin{figure}
\epsfxsize=11truecm
\hspace{2.75truecm}
\epsfbox{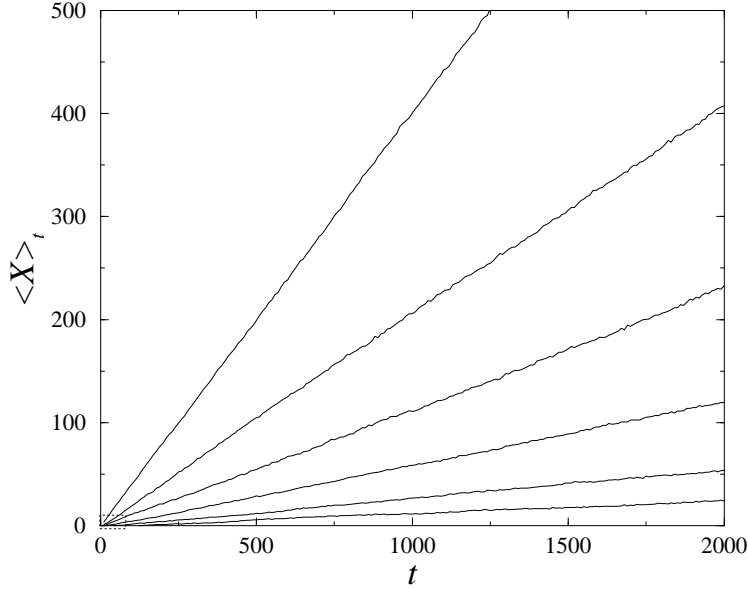}
\caption{
The average position of the piston as a function of time 
for the infinite system. The parameters are 
$m=1$, $T_{[1]}=1$, $T_{[2]}=10$, $n_{[1]}=1$, $n_{[2]}=1/10$ and 
$M=2,5,10,20,50,100$ (from top to bottom). The dotted box is enlarged
on Fig. \ref{fig2b}.
\label{fig2}}
\end{figure}

\begin{figure}
\epsfxsize=11truecm
\hspace{2.75truecm}
\epsfbox{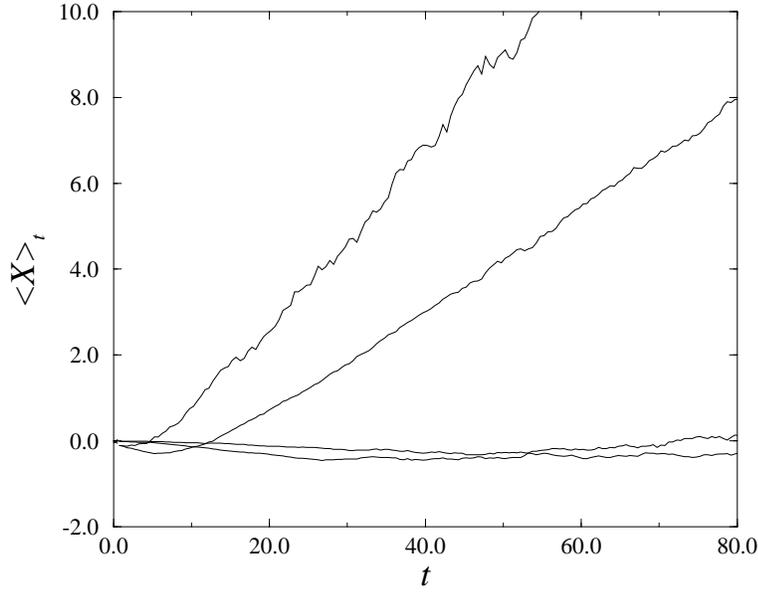}
\caption{
The average position of the piston as a function of time 
for the infinite system. Magnification of the dotted box on Fig. \ref{fig2}. 
The parameters are as in Fig. \ref{fig2} with
$M=5,10,50,100$.  
\label{fig2b}}
\end{figure}

For the infinite system, we simulate the openness of the system 
with  sources of in-going particles very far from the piston position. 
We compute the average time dependent position $\langle X\rangle_t$ 
of the piston on $10^3-10^4$ different samples. 
An example of the time evolution of the piston average position 
is shown on Fig. \ref{fig2} for the following parameter
$m=1$, $T_{[1]}=1$, $T_{[2]}=10$, $n_{[1]}=1$, $n_{[2]}=1/10$, {\sl i.e.}
equal pressure on both sides of the piston, 
and $M=2,5,10,20,50,100$ (from top to bottom).
The initial position and velocity of the piston are
$X_0=V_0=0$. As expected from Eq.(\ref{139}) 
we observe that the average position quickly behaves
as $\langle X\rangle_t=\lla V\rra_\infty t$ ($t\gtrsim 10^2$). 
The relaxation time $\tau_a$ necessary 
to reach this stationary behavior is too short to be represented on
Fig. \ref{fig2}, and is presented on Fig. \ref{fig2b} for $M=5,10,50,100$. 
Remark that $\tau_a$ depends on the ratio $m/M$.

\begin{figure}
\epsfxsize=11truecm
\hspace{2.75truecm}
\epsfbox{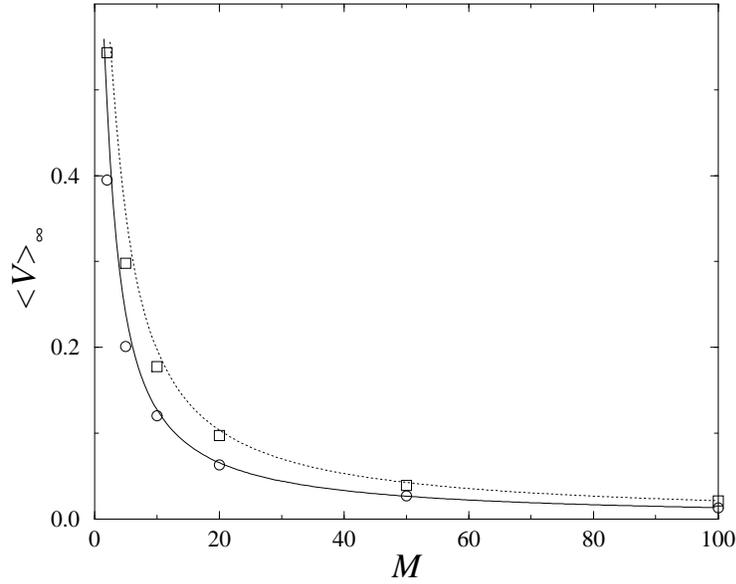}
\caption{
The stationary average velocity reached by the piston for the infinite system
as a function of its mass $M$.
The parameters of the simulation are $m=1$, $T_{[1]}=1$, $n_{[1]}=1$ while
$T_{[2]}=10$, $n_{[2]}=1/10$ (circles) and $T_{[2]}=20$, $n_{[2]}=1/20$ 
(squares), respectively. The solid, resp. dashed, 
line shows the theoretical value obtained in the text up to fourth 
order in $\ee$.  
\label{fig3}}
\end{figure}

\noi
Even if the pressures 
are equal on both sides of the piston, 
{\sl i.e.} $n_{[1]}T_{[1]}=n_{[2]}T_{[2]}$, the stationary velocity
$\lla V\rra_\infty$ is non zero
and is presented as a function of $M$ on 
Fig. \ref{fig3} for the above parameters, as well as 
for $T_{[2]}=20$ and $n_{[2]}=1/20$.
The lines on Fig. \ref{fig3} show the average stationary velocity computed
in Sec.\ref{6.2} up to the fourth order in $\ee=\sqrt{m/M}$
\beq
\lla V\rra_\infty=\ee\lla V\rra_\infty^{(1)}+\ee^3\lla V\rra_\infty^{(3)}
+{\cal O}(\ee^5)
\eeq
with (see Eq.(\ref{102}))
\beq
\lla V\rra_\infty^{(1)}
={\sqrt{2\pi}\over 4}{1\over \sqrt{M}}\left(\sqrt{T_{[2]}}
-\sqrt{T_{[1]}}\right)
\eeq
and
\beq
\lla V\rra_\infty^{(3)}
={\sqrt{2\pi}\over 96}{1\over \sqrt{M}}\left[(16-3\pi){\left(\sqrt{T_{[2]}}
-\sqrt{T_{[1]}}\right))^3\over 
\sqrt{T_{[1]}T_{[2]}}}-6\left(\sqrt{T_{[2]}}-\sqrt{T_{[1]}}\right)\right].
\eeq
One sees that the values obtained at the fourth order agree very well  
with the numerical results.

\begin{figure}
\epsfxsize=11truecm
\hspace{2.75truecm}
\epsfbox{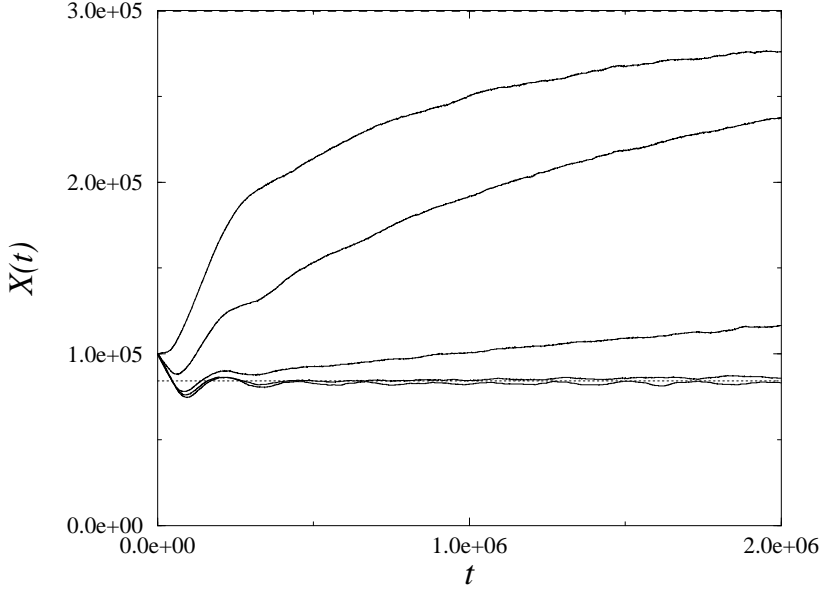}
\caption{
The piston's position as a function of time. 
The parameters of the simulation are $m=1$, $T_{[1]}=1$, $n_{[1]}=1$ while
$T_{[2]}=10$, $n_{[2]}=1/5$. $X(t)$ is shown for
for $M=2,10,10^2,10^3,10^4$ (from top to bottom).
The dashed line stands for the equilibrium solution 
($X=3\times 10^5$) while the dotted lines show the 
thermodynamical prediction ($X=84260.5$) for the adiabatic piston 
($\kappa=0$).
\label{fig4}}
\end{figure}

Concerning the finite system, we present here a qualitative study of the
piston properties as a function of the parameter $\ee$. 
We consider a finite system of size $L=6\times 10^5$ 
with $N_{[1]}=N_{[2]}=10^5$ and with
$X_0=10^5$ (thus we have $n_{[1]}=1$, $n_{[2]}=1/5$). 
The initial temperatures
on the left and on the right of the piston are $T_{[1]}=1$ and $T_{[2]}=10$.  
The initial pressure difference is 
thus $p_{[1]}-p_{[2]}=k_B(n_{[1]}T_{[1]}-n_{[2]}T_{[2]})<0$, 
pushing the piston toward the left side. 
The time scale of our simulations has to be compared with the typical time
it takes for a given particle to cross the system which is of the order 
${\cal O}(10^5)$.
We plot on Fig. \ref{fig4} the
time behavior of the piston's position for different values of 
$\ee=2^{-1/2},10^{-n/2}$ ($n=1,2,3,4$). 

\begin{figure}
\epsfxsize=11truecm
\hspace{2.75truecm}
\epsfbox{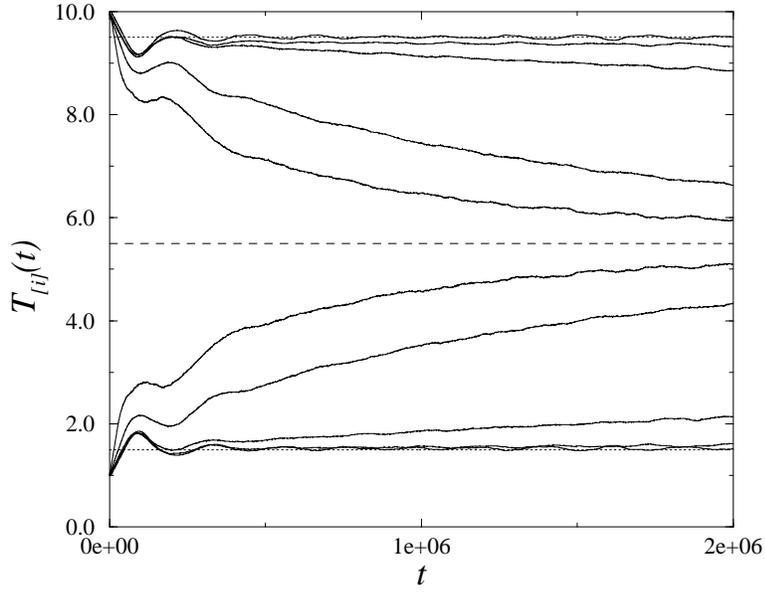}
\caption{
The temperature of the fluids on both sides of the piston 
as a function of time. 
The parameters of the simulation are $m=1$, $T_{[1]}=1$, $n_{[1]}=1$ while
$T_{[2]}=10$, $n_{[2]}=1/5$. $T_{[1]}(t)$, resp. $T_{[2]}(t)$, is shown for
for $M=2,10,10^2,10^3,10^4$ (from middle to bottom, resp. to top).
The dashed line stands for the equilibrium solution 
($T_{[1]}(\infty)=T_{[2]}(\infty)=(T_{[1]}+T_{[2]})/2=5.5$) while the 
dotted lines show the 
thermodynamical prediction ($T_{[1]}(\infty)\sim 1.545 $, 
$T_{[2]}(\infty)\sim 9.455$) for $\kappa=0$.
\label{fig5}}
\end{figure}

\begin{figure}
\epsfxsize=11truecm
\hspace{2.75truecm}
\epsfbox{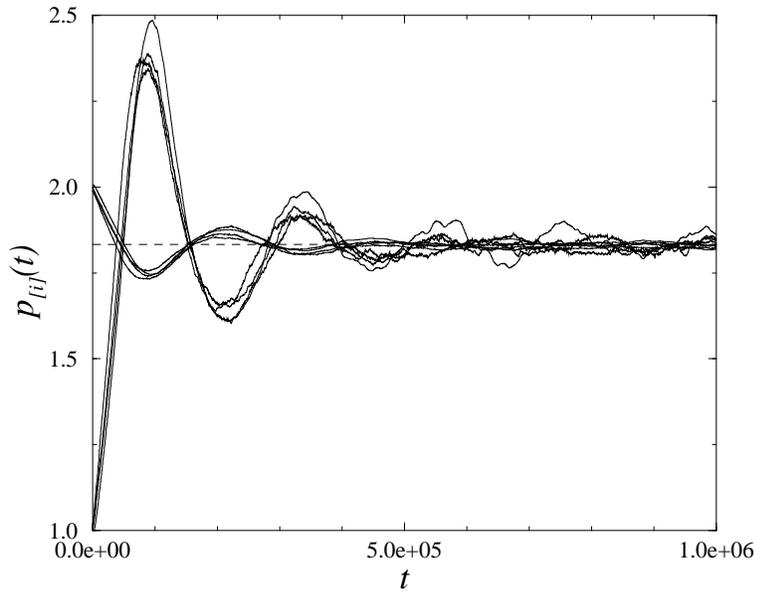}
\caption{
The pressures of the fluids on both sides of the piston 
as a function of time. 
The parameters of the simulation are $m=1$, $T_{[1]}=1$, $n_{[1]}=1$, 
$p_{[1]}=1$  while
$T_{[2]}=10$, $n_{[2]}=1/5$, $p_{[2]}=2$. 
$p_{[1]}(t)$, resp. $p_{[2]}(t)$, is shown for
for $M=10,10^2,10^3,10^4$.
The dashed line stands for the equilibrium solution which is equal to
the thermodynamical  prediction 
($p_{[1]}(\infty)=p_{[2]}(\infty)=N_{[1]}(T_{[1]}+T_{[2]})/L=11/6$).
\label{fig6}}
\end{figure}

\noi
On Fig. \ref{fig5} and Fig. \ref{fig6}, we plotted 
respectively the evolution of the temperatures $T_{[i]}(t)$, 
$(i=1,2)$ and of the pressures 
for the same parameters.
Notice that the temperatures are estimated through the second moment
of the fluid particles velocity for all the particles present on the same 
side of the piston. It is equivalent to the temperatures on the 
face of the piston only if we suppose that the system is homogeneous.
The dashed lines show the equilibrium solution predicted in 
Sec.\ref{equilfinite} where
$n_{[1]}(\infty)=n_{[2]}(\infty)=(N_{[1]}+N_{[2]})/L$,
$X_{\infty}=LN_{[1]}/(N_{[1]}+N{[2]})$, $T_{[1]}(\infty)=T_{[2]}(\infty)=
(T_{[1]}+T_{[2]})/2$, $p_{[1]}(\infty)=p_{[2]}(\infty)=k_B (N_{[1]}+N_{[2]})
(T_{[1]}+T_{[2]})/(2L)$. 
On the other hand, the dotted lines show the equilibrium values obtained 
from a numerical integration of  
the phenomenological thermodynamical equations (\ref{eqv},\ref{42},\ref{43})
with $\kappa=0$ for the primitive model of an adiabatic piston and  
where we used $\lll_{[i]}(V)\simeq 
\lll_{[i]}(V=0)=n_{[i]}(t)\sqrt{8k_B T_{[i]} m/\pi}$. In this case the system
evolves toward a metastable state with $p_{[1]}(\infty)=p_{[2]}(\infty)$ but
$T_{[1]}(\infty)\ne T_{[2]}(\infty)$. One should note that on the scale 
of Figs. \ref{fig4}, \ref{fig5} and \ref{fig6}, one can hardly distinguish
the results obtained by  numerical integration ($X\simeq 84260.5$, 
$T_{[1]}\simeq 1.545$, $T_{[2]}\simeq 9.455$ and $p_{[1]}=p_{[2]}\simeq 1.83$) 
from those obtained
by heuristic arguments Eqs.(\ref{151}-\ref{154}),
$X\simeq 82200$, 
$T{[1]}\simeq 1.48$, $T_{[2]}\simeq 9.32$ and $p_{[1]}=p_{[2]}\simeq 1.80$.

\noi
Let us first comment on the Fig. \ref{fig6} where one sees that the
pressures quickly become constant and equal to their final values.
The time scale $t_a$ needed to equalize the pressures is roughly independent 
on the mass ratio $m/M$. This behavior has been used 
in Sec.\ref{7.2} to give a qualitative discussion of the evolution. 
Consider now the Fig. \ref{fig5} where one sees that the temperatures reach
quickly an intermediate stage characterized by the equality of the pressures 
(time scale $t_a$) then evolve slowly toward their equilibrium values on 
a time scale $\tau_{\rm e}$. The time scale $\tau_{\rm e}$ 
depends strongly on the mass ratio $\ee^2=m/M$ and for 
$\ee\ll 1$ we have $\tau_{\rm e}\gg t_a$. One remarks that, for times
covered in  our simulations and
for $\ee^2\simeq 10^{-4}$ the system behaves as if the
piston was ``truly'' adiabatic ($\kappa=0$).

\begin{figure}
\epsfxsize=11truecm
\hspace{2.75truecm}
\epsfbox{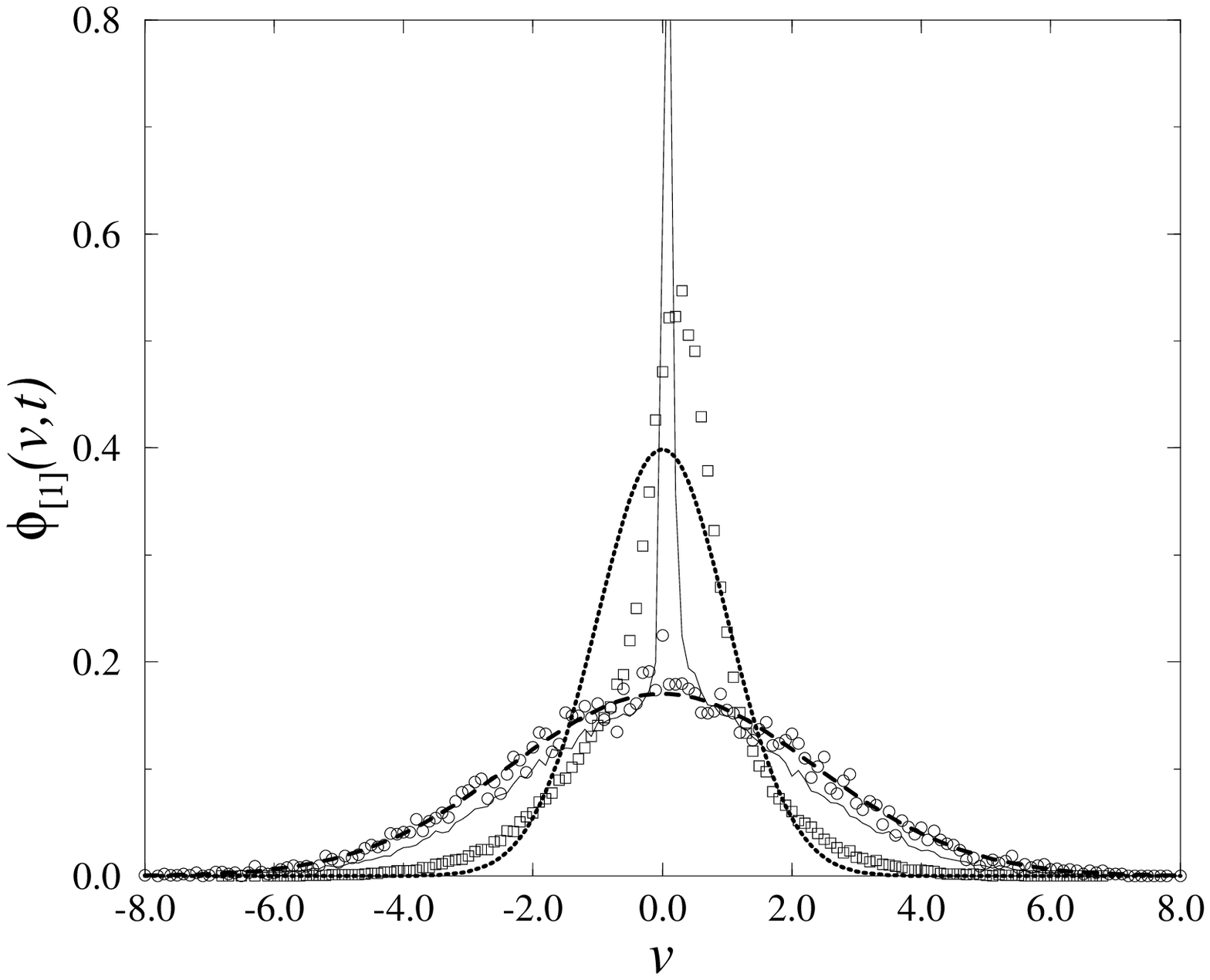}
\caption{
The particles velocity distribution $\phi_{[1]}(v,t)$ 
of the fluid $[1]$ at different time for the finite system.
The initial conditions are 
$T_{[1]}=1$, $T_{[2]}=10$, $p_{[1]}=p_{[2]}=1$, $X_0=V_0=0$, $m=1$ and $M=10$.
The dotted line show the initial distribution $\phi_{T_{[1]}}(v)$ while the
dashed line show the expected equilibrium distribution 
$\phi_{T_{[1]}(\infty)}(v)$ with $T_{[1]}(\infty)=(T_{[1]}+T_{[2]})/2$.
The squares show the velocity distribution for $t_1=2\times 10^5$ where the 
temperature of the fluid is $T_{[1]}(t_1)\simeq  1.6$. 
The solid line is for $t_2=2.5\times 10^6$ with $T_{[1]}(t_2)\simeq 3.7$ while 
the circles are for $t_3=7.5\times 10^7$ where 
$T_{[1]}(t_3)\simeq T_{[1]}(\infty)$.
\label{fig7}}
\end{figure}

\begin{figure}
\epsfxsize=11truecm
\hspace{2.75truecm}
\epsfbox{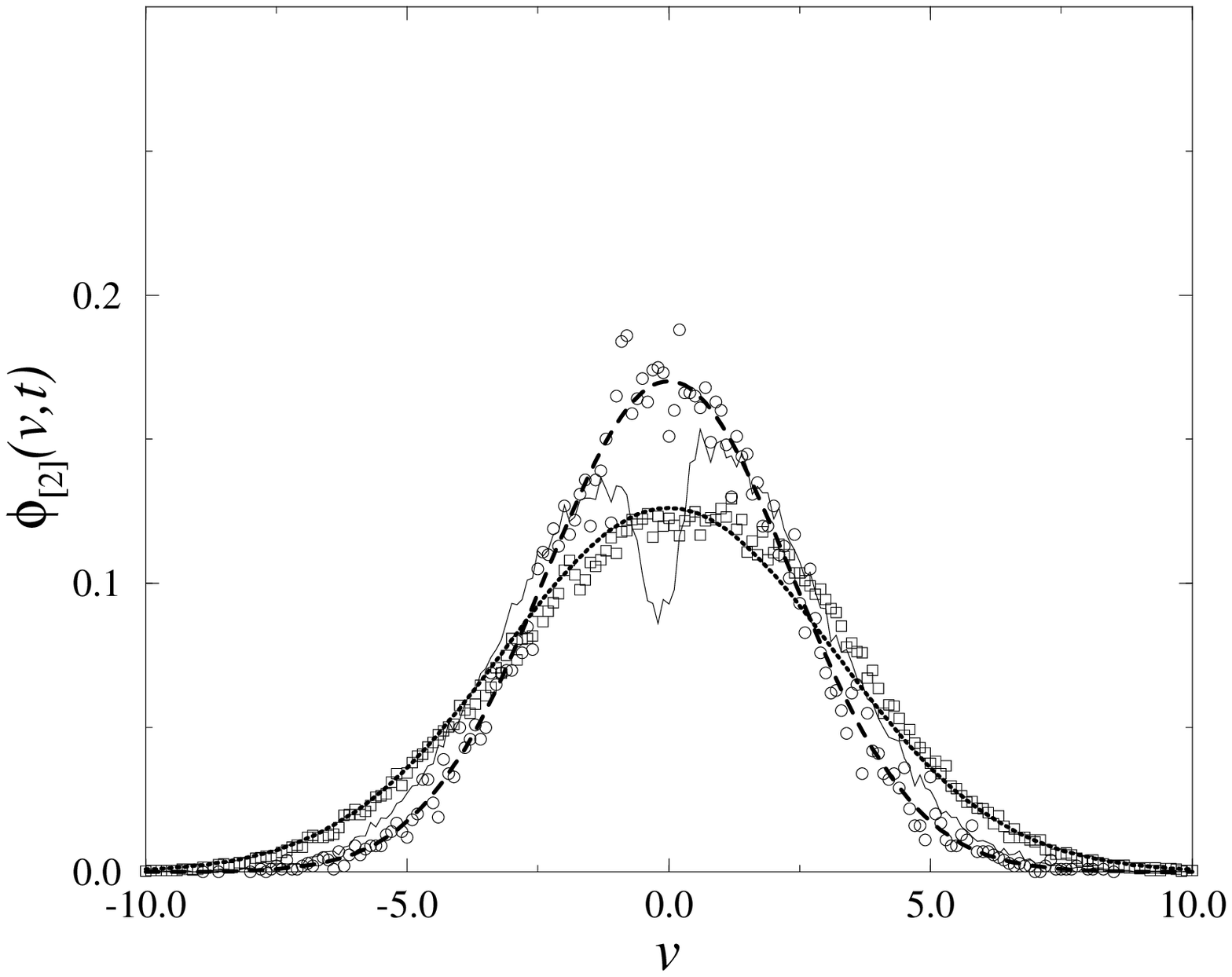}
\caption{
The particles velocity distribution $\phi_{[2]}(v,t)$ 
of the fluid $[2]$ at different time for the finite system.
The initial conditions are 
$T_{[1]}=1$, $T_{[2]}=10$, $p_{[1]}=p_{[2]}=1$, $X_0=V_0=0$, $m=1$ and $M=10$.
The dotted line show the initial distribution $\phi_{T_{[2]}}(v)$ while the
dashed line show the expected equilibrium distribution 
$\phi_{T_{[2]}(\infty)}(v)$ with $T_{[2]}(\infty)=(T_{[1]}+T_{[2]})/2$.
The squares show the velocity distribution for $t_1=2\times 10^5$ where the 
temperature of the fluid is $T_{[2]}(t_1)\simeq  9.4$. 
The solid line is for $t_2=2.5\times 10^6$ with $T_{[2]}(t_2)\simeq 7.3$ while 
the circles are for $t_3=7.5\times 10^7$ 
where $T_{[2]}(t_3)\simeq T_{[2]}(\infty)$.
\label{fig8}}
\end{figure}

We plotted on Fig. \ref{fig7} and Fig. \ref{fig8} the velocity distribution 
for the particles of both the fluids. It is averaged on all 
particles located on the same side of the piston. The initial conditions are 
$T_{[1]}=1$, $T_{[2]}=10$, $p_{[1]}=p_{[2]}=1$, 
$X_0=V_0=0$, $m=1$ and $M=10$.
The dotted lines show the Maxwellian 
initial distribution $\phi_{T_{[i]}}(v)$.
We present these distributions for three different times 
$t=t_1,t_2,t_3$. 

\begin{figure}
\epsfxsize=11truecm
\hspace{2.75truecm}
\epsfbox{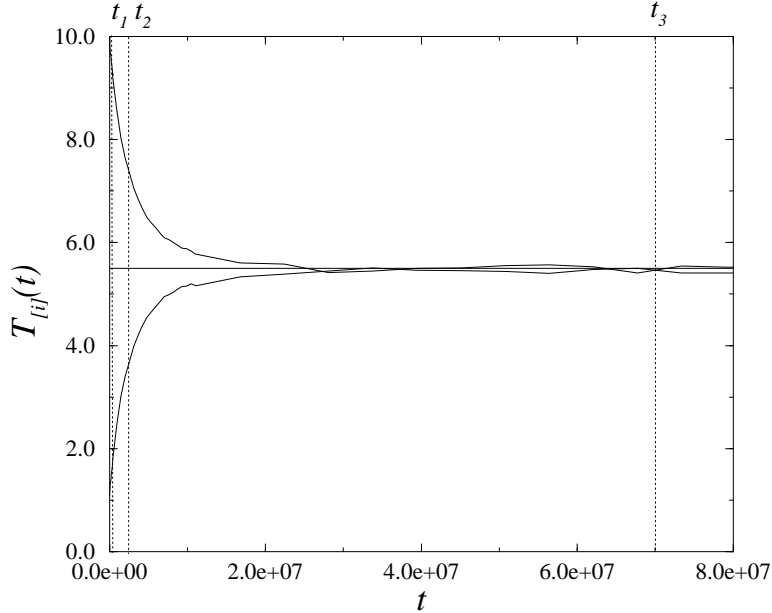}
\caption{
The time evolution of the 
temperatures $T_{[1]}(t)$ (bottom) and $T_{[2]}(t)$ (up) for the same 
initial conditions as for Figs. \ref{fig7} and \ref{fig8}. 
The successive vertical dotted  
lines show the times $t_1,t_2$ and $t_3$ at which 
the velocity distributions plotted on 
Figs. \ref{fig7} and \ref{fig8} were determined.
\label{fig9}}
\end{figure}

\noi
For the time $t_1=2\times 10^5$ the system has not reached the relaxation time
$t_a$ and the pressures are not yet constant. 
At the time $t_2$ ($t_a<t_2<\tau_{\rm e}$), the system has constant pressures 
but has not reached the equilibrium state.
At time $t_3>\tau_{\rm e}$  the system has reached its final 
equilibrium state. We present on Fig. \ref{fig9} the time dependence of the
fluid temperatures. Vertical lines stand for times $t_1,t_2,t_3$.
We see that, for times such as $t<\tau_{\rm e}$, 
these distributions are not Maxwellian 
but develops peaks or wells for small velocities.
Moreover these intermediate time distribution are clearly not
symmetric.
This facts are the reason 
why we did not suppose Maxwellian distributions for the fluid particles in our 
previous analytical study. After a sufficiently long time of the
order of the equilibrium relaxation time of the system $\tau_e$, 
the distributions are very well fitted with the Maxwellian distribution 
and we conjecture that $\phi_{[1]}(v,\infty)=\phi_{[2]}(v,\infty)=
\phi_{T_{[1]}(\infty)}(v)$ with $T_{[1]}(\infty)= (T_{[1]}+T_{[2]})/2=11/2$.
The dashed lines on Fig. \ref{fig7} and Fig. \ref{fig8} show the 
conjectured equilibrium distribution.
This last conjecture is valid only for $m<M$ as it has been shown that the 
equilibrium distribution in the case of $m=M$ is a simple superposition 
of the two initial Maxwellian \cite{sinai}.

\section{Conclusions}\label{conclusion}

Several conclusions can be drawn from this investigation on the stochastic
motion of a piston which is adiabatic when rigidly fixed.

For infinite systems, it has been shown that very rapidly a stationary
state is reached, where the average velocity of the piston is a function of the
pressures and temperatures of the fluids on both sides.
In particular if the pressures are equal but the temperatures different 
the final average velocity is non-zero and the piston has a 
macroscopic motion toward the high temperature region.
This result is related to the asymmetry in the fluctuations of
the piston, due to the asymmetry of the temperatures on both sides, 
which in turn induces both a macroscopic force and a heat 
current from one side to the other. In other words in the stationary state 
the stochastic piston is a conductor. To obtain a stationary state with zero
average velocity it is necessary to compensate the stochastic force
by a macroscopic difference in the pressures of the fluids. 
In this very special state no work is done by the piston on the fluids, 
but heat is transferred from one side to the other.
Moreover this state is not a state of mechanical equilibrium since
all the odd moments of the piston velocity, except the first one, 
are non zero.

For finite systems it has been shown that the system will always evolve 
toward the equilibrium state predicted for a conducting piston where
pressures, temperatures and densities are equal on both 
sides of the piston , but the time
needed to reach this final equilibrium state will depend strongly on the 
mass ratio $m/M$ and can reach several time the age of the universe for
reasonable numbers. This evolution takes place in two stages. 
In the first stage the system reaches rather rapidly (although not as fast as 
the time needed to reach the stationary state of the infinite system) a state 
of ``metastable equilibrium'' where the pressures are equal 
but the temperatures different. This initial evolution corresponds 
to an adiabatic evolution (no heat transfer) and the results obtained
from the microscopical theory, from the thermodynamic equations, and from
the numerical simulations are in good agreement.
In the second stage the piston and the temperatures of the fluids 
evolve in such a way that the pressures remain approximatively constant 
and equal. If $m\ll M$ it appears that at all time the average velocity
of the piston co\"\i ncide with the velocity in the stationary state
 of the infinite system with values of temperatures and pressures given by 
those of the finite system at that time. This observation can be understood
from the fact that the stationary state of the infinite system
is reached very rapidly. If $m\ll M$ the time needed to reach the 
final state is so large that on the numerical simulations the 
``metastable equilibrium'' state appear to be stable and the piston behaves 
as an adiabatic piston on the time scale considered. 
On the other hand for $m\simeq M$ the final equilibrium state is reached rather
rapidly. Moreover it was observed that during the time evolution the
velocity distribution of the fluid particles is not Maxwellian and that
this distribution evolves slowly toward the equilibrium 
Maxwellian distribution characterized by the equilibrium temperature. 

Finally the conclusions obtained from thermodynamics, from kinetic theory,
and from numerical simulations, are all in good agreements.
There appears to be no violation of the second law and no
paradox involved.

\acknowledgements

One of the authors (Ch. G.) is very much indebted to J. L. Lebowitz and
E. Lieb for stimulating and continuous e-mail discussions and remarks, 
which have been used for the writing of the Introduction.

\end{document}